\documentclass[useAMS,usenatbib]{mn2e}
\input psfig.sty

\def \aj {AJ}
\def \apj {ApJ}
\def \apjl {ApJL}
\def \mnras {MNRAS}
\def \apjs {ApJS}
\def \aap {A\&A}
\def \etal {et~al.~}

\def \spose#1{\hbox  to 0pt{#1\hss}}  
\def \lta{\mathrel{\spose{\lower 3pt\hbox{$\sim$}}\raise  2.0pt\hbox{$<$}}}
\def \gta{\mathrel{\spose{\lower  3pt\hbox{$\sim$}}\raise 2.0pt\hbox{$>$}}}
\def \ion#1#2{#1{\footnotesize{#2}}\relax} 
\newcommand {\ha}  {\ifmmode H\alpha \else H$\alpha $ \fi} 
\newcommand {\hi} {\ion{H}{I}} 
\newcommand {\po} {\ifmmode \phantom{0} \else $\phantom{0}$ \fi}
\newcommand {\pt} {\ifmmode \phantom{2} \else $\phantom{0}$ \fi}

\def \kms {\ifmmode  \,\rm km\,s^{-1} \else $\,\rm km\,s^{-1}  $ \fi }
\def \kpc {\ifmmode  {\rm kpc}  \else ${\rm  kpc}$ \fi  }  
\def \hkpc {\ifmmode  {h^{-1}\rm kpc}  \else ${h^{-1}\rm kpc}$ \fi  }  
\def \hMpc {\ifmmode  {h^{-1}\rm Mpc}  \else ${h^{-1}\rm Mpc}$ \fi  }
\def \Lsun {\ifmmode {\rm L_{\odot}} \else ${\rm L_{\odot}}$ \fi}   
\def \Msun {\ifmmode {\rm M_{\odot}} \else ${\rm M_{\odot}}$ \fi} 
\def \hMsun {\ifmmode {h^{-1}\,\rm M_{\odot}} \else ${h^{-1}\,\rm M_{\odot}}$ \fi}
\def \hhMsun {\ifmmode {h^{-2}\,\rm M_{\odot}}\else ${h^{-2}\,\rm M_{\odot}}$ \fi}
\def \Mstar {\ifmmode M_{\rm star} \else $M_{\rm star}$ \fi} 

\def\LCDM{$\Lambda$CDM }
\def \LCDM {\ifmmode \Lambda{\rm CDM} \else $\Lambda{\rm CDM}$ \fi}
\def \sig8 {\ifmmode \sigma_8 \else $\sigma_8$ \fi} 
\def \OmegaM {\ifmmode \Omega_{\rm M} \else $\Omega_{\rm M}$ \fi} 
\def \Omegab {\ifmmode \Omega_{\rm b} \else $\Omega_{\rm b}$ \fi} 
\def \OmegaL {\ifmmode \Omega_{\rm \Lambda} \else $\Omega_{\rm \Lambda}$\fi} 
\def \Deltavir {\ifmmode \Delta_{\rm vir} \else $\Delta_{\rm vir}$ \fi}
\def \rhocrit {\ifmmode \rho_{\rm crit} \else $\rho_{\rm crit}$ \fi}

\def \rs {\ifmmode r_{\rm s} \else $r_{\rm s}$ \fi} 
\def \Rvir {\ifmmode R_{\rm vir} \else $R_{\rm vir}$ \fi}
\def \Vvir {\ifmmode V_{\rm  vir} \else  $V_{\rm vir}$  \fi} 
\def \Vmax {\ifmmode V_{\rm  max} \else  $V_{\rm max}$  \fi} 
\def \Vmaxh {\ifmmode V_{\rm  max,h} \else  $V_{\rm max,h}$  \fi} 
\def \Mvir {\ifmmode M_{\rm  vir} \else $M_{\rm  vir}$ \fi}  
\def \Nvir {\ifmmode N_{\rm  vir} \else $N_{\rm  vir}$ \fi}  
\def \Jvir {\ifmmode J_{\rm vir} \else $J_{\rm vir}$ \fi} 
\def \Evir {\ifmmode E_{\rm vir} \else $E_{\rm vir}$ \fi} 
\def \lam {\ifmmode \lambda  \else $\lambda$ \fi} 
\def \lamp {\ifmmode \lambda^{\prime} \else $\lambda^{\prime}$  \fi} 
\def \lampc {\ifmmode \lambda^{\prime}_{\rm c} \else
  $\lambda^{\prime}_{\rm c}$  \fi} 

\def \xoff {\ifmmode x_{\rm off} \else $x_{\rm off}$ \fi}
\def \rhorms {\ifmmode \rho_{\rm rms} \else $\rho_{\rm rms}$ \fi}
\def \qbar {\ifmmode \bar{q} \else $\bar{q}$ \fi}

\def \mstar {\ifmmode m_{\rm star} \else $m_{\rm star}$ \fi} 
\def \mgal {\ifmmode m_{\rm gal} \else $m_{\rm gal}$ \fi} 
\def \lamgal {\ifmmode \lambda_{\rm gal} \else $\lambda_{\rm gal}$ \fi} 
\def \Vcirc {\ifmmode V_{\rm circ} \else $V_{\rm circ}$ \fi} 
\def \Vobs {\ifmmode V_{\rm obs} \else $V_{\rm obs}$ \fi} 
\def \Vopt {\ifmmode V_{\rm opt} \else $V_{\rm opt}$ \fi} 
\def \YB {\ifmmode \Upsilon_B \else $\Upsilon_B$ \fi} 
\def \DeltaIMF {\ifmmode \Delta_{\rm IMF} \else $\Delta_{\rm IMF}$ \fi}
\def \fracdev {\ifmmode {frac\_{deV}} \else $frac\_{deV}$ \fi}

\title[Connecting Galaxies with Dark Matter Haloes]{The Kinematic
  Connection Between Galaxies and Dark Matter Haloes}

\author[Dutton et al.]  {Aaron  A.
  Dutton$^{1,2}$\thanks{dutton@uvic.ca}\thanks{CITA National Fellow}, 
  Charlie Conroy$^3$, Frank C. van den Bosch$^4$,
 \newauthor {Francisco Prada$^5$ \& Surhud More$^6$}\\ 
  $^1$UCO/Lick Observatory, University of California, Santa Cruz, CA 95064, USA\\ 
  $^2$Dept. of Physics and Astronomy, University of Victoria, Victoria, BC, V8P 5C2, Canada\\
  $^3$Department of Astrophysical Sciences, Peyton Hall, Princeton University, Princeton, NJ 08544\\
  $^4$Department of Physics and Astronomy, University of Utah, 115 South 1400 East, Salt Lake City, UT 84112-0830\\
  $^5$Instituto de Astrofisica de Andalucia (CSIC), E18008 Granada, Spain\\
  $^6$Kavli Institute for Cosmological Physics, University of Chicago, 933 East
56th Street, Chicago, IL 60637, USA}

\begin{document}
             
\date{accepted to MNRAS}
             
\pagerange{\pageref{firstpage}--\pageref{lastpage}}\pubyear{2010}

\maketitle           

\label{firstpage}
             

\begin{abstract}
  Using estimates of dark halo masses from satellite kinematics, weak
  gravitational lensing, and halo abundance matching, combined with
  the Tully-Fisher and Faber-Jackson relations, we derive the mean
  relation between the optical, $\Vopt$, and virial, $V_{200}$,
  circular velocities of early- and late-type galaxies at redshift
  $z\simeq 0$.  For late-type galaxies $\Vopt\simeq V_{200}$ over the
  velocity range $\Vopt = 90-260 \kms$, and is consistent with $\Vopt
  = \Vmaxh$ (the maximum circular velocity of NFW dark matter haloes
  in the concordance \LCDM cosmology). However, for early-type
  galaxies $\Vopt \ne V_{200}$, with the exception of early-type
  galaxies with $\Vopt \simeq 350 \kms$.  This is inconsistent with
  early-type galaxies being, in general, globally isothermal.  For low
  mass ($\Vopt \lta 250 \kms$) early-types $\Vopt > \Vmaxh$,
  indicating that baryons have modified the potential well, while high
  mass ($\Vopt \gta 400 \kms$) early-types have $\Vopt < \Vmaxh$.
  Folding in measurements of the black hole mass - velocity dispersion
  relation, our results imply that the supermassive black hole - halo
  mass relation has a logarithmic slope which varies from $\simeq 1.4$
  at halo masses of $\simeq 10^{12}h^{-1}\Msun$ to $\simeq 0.65$ at
  halo masses of $10^{13.5}h^{-1}\Msun$.  The values of
  $\Vopt/V_{200}$ we infer for the Milky Way and M31 are lower than
  the values currently favored by direct observations and dynamical
  models. This offset is due to the fact that the Milky Way and M31
  have higher $\Vopt$ and lower $V_{200}$ compared to typical
  late-type galaxies of the same stellar masses. We show that current
  high resolution cosmological hydrodynamical simulations are unable
  to form galaxies which simultaneously reproduce both the
  $\Vopt/V_{200}$ ratio and the $\Vopt-\Mstar$
  (Tully-Fisher/Faber-Jackson) relation.
 \end{abstract}


\setcounter{footnote}{1}


\section{Introduction}
\label{sec:intro}

It is theoretically expected and observationally established that
galaxies are surrounded by extended haloes of dark matter (White \&
Rees 1978; Blumenthal \etal 1984; van Albada \& Sancisi 1986; Zaritsky
\& White 1994; Brainerd \etal 1996; Prada \etal 2003).  In recent
years much progress has been made in understanding the relation
between the masses of dark matter haloes and the properties of the
galaxies that reside in them.  The majority of work in the literature
has focused on the relation between halo mass and galaxy luminosity or
stellar mass (e.g.  Yang \etal 2003; Kravtsov \etal 2004; van den
Bosch \etal 2004; Hoekstra \etal 2005; Mandelbaum \etal 2006; Conroy
\etal 2007; More \etal 2009).  An alternative approach is to link
galaxies to haloes via kinematics, for example, by measuring the
relation between the circular velocity ($\Vcirc(r)= \sqrt{G M
  (<r)/r}$, for a spherical system) within the optical part (e.g. the
half light radius or 2.2 disk scale lengths) of galaxies, $\Vopt$, to
the circular velocity at the virial radius of the dark matter halo,
$V_{200}$.

This approach has the advantage that it is free from uncertainties in
luminosities or stellar masses due to uncertainties in extinction from
dust, stellar populations or the stellar initial mass function
(IMF). In addition, because it is a {\it dynamical} link, it gives a
direct measurement of the slope of the global mass density profile
within the virial radius. For example, for a singular isothermal
density profile, $\rho(r) \propto r^{-2}$, and thus $V_{\rm c}
=const.$ which implies $\Vopt/V_{200}=1$.  By measuring a departure
from this prediction, one can rule out galaxies being globally
isothermal. This is a relevant issue because it is common practice to
infer halo masses from observed $\Vopt$ by assuming $\Vopt=V_{200}$,
even though this assumption lacks both theoretical and observational
support. For example, in \LCDM cosmologies dark matter haloes are not
isothermal, so a detection of non-isothermality would be useful from a
\LCDM haloes perspective.

The $\Vopt/V_{200}$ ratio also contains information on the relative
importance of baryons vs dark matter in the optical regions of
galaxies. For galaxy mass \LCDM haloes, the ratio between the maximum
circular velocity of the dark matter halo, $\Vmaxh$, and the virial
velocity of the dark matter halo $V_{\rm max,h}/V_{200}\simeq
1.1-1.2$ (Bullock \etal 2001), and thus if $\Vopt/V_{200}$ is
observed to be significantly larger than this, it indicates that
baryons have modified the potential well.  Using weak lensing
measurements of halo masses combined with the Tully-Fisher (1977,
hereafter TF) relation, Seljak (2002) inferred $\Vopt/V_{200}=1.8$
(with a 95\% C.I. lower limit of 1.4), for late-type $L_*$
galaxies. Such high values of $\Vopt/V_{200}$ are naturally explained
by the combined effects of baryons adding to the optical circular
velocity directly, and indirectly by inducing contraction of the dark
matter halo (Blumenthal \etal 1986).

However, a high value of $\Vopt/V_{200}$ introduces problems in
reconciling the TF relation with the halo mass function and galaxy
luminosity function, LF. Semi-analytic galaxy formation models that
are able to simultaneously reproduce the zero point of the TF relation
and the galaxy LF assume that $\Vopt = V_{200}$ (e.g.  Somerville \&
Primack 1999), or $\Vopt=V_{\rm max,h}$ (e.g.  Croton \etal
2006). However, models that take into account the contribution of the
baryons to the rotation curve and the effect of halo contraction have
been unable to match both the LF and TF relation (e.g. Benson \etal
2003; Benson \& Bower 2010).  Dutton \etal (2007) showed that the
$\Vopt/V_{200}$ ratio can place strong constraints on dark halo
structure. In particular they argued that models with adiabatic
contraction, standard $\Lambda$CDM halo concentrations, and standard
stellar IMFs are unable to {\it simultaneously} reproduce the zero
points of the TF and size-luminosity relations, and the requirement
that $\Vopt\simeq \Vmaxh$ (in order to reproduce the LF).  However, if
halo contraction was somehow avoided, or even if haloes could {\it
  expand} in response to galaxy formation, then models could be
constructed that reproduced all of the observational constraints.

In this paper we derive the relation between $\Vopt$ and $V_{200}$ for
both early- (red, bulge dominated) and late-type (blue disk dominated)
galaxies at redshifts $z\simeq 0$. For late-types, we use
$\Vopt=V_{2.2}$, where $V_{2.2}$ is the rotation velocity at 2.2 disk
scale lengths. For early-types we use $\Vopt=1.65\sigma(R_{50})$,
where $\sigma(R_{50})$ is the velocity dispersion within the projected
half light radius of the galaxy, and the factor of 1.65 is from
Padmanabhan \etal (2004, see also \S\ref{sec:fj}). We determine the
mean relation between virial velocity and stellar mass using published
measurements of halo masses from satellite kinematics, weak lensing,
and halo abundance matching studies. By comparing these relations with
the observed TF and Faber Jackson (1976, hereafter FJ) relations we
derive the mean relation between $\Vopt$ and $V_{200}$. In a future
paper (Dutton \etal in prep) we will combine these results with other
scaling relations of early- and late-type galaxies to place
constraints on the structure of their dark matter haloes.

This paper is organized as follows.  In \S2 we describe the
observations of halo masses and galaxy scaling relations.  In \S3 we
derive and discuss the relation between $\Vopt$ and $V_{200}$.  In \S4
we discuss the implications of our results for the slope of the
relation between super-massive black hole mass and halo mass.  In \S5
we compare our results with predictions for $\Lambda$CDM dark matter
haloes. In \S6 we compare our results with halo masses derived for the
Milky Way and M31 galaxies, as well as galaxies formed in cosmological
hydrodynamical simulations. In \S7 we give a summary.

Unless otherwise specified, throughout this paper we adopt a Hubble
parameter, $H_0=100\, h \,\rm km \,s^{-1}\, Mpc^{-1}$, i.e. stellar
masses are expressed in $h^{-2}\Msun$ units, halo masses are expressed
in $h^{-1}\Msun$ units, and halo sizes are expressed in $h^{-1} \kpc$
units.

\begin{table*}
 \caption{Cosmological and stellar mass parameters adopted by
   studies of the halo masses - stellar mass relation.
Notes--- Col. (2) Method used to derive the halo mass - stellar mass relation: GC--Group Catalog; 
AM--Abundance Matching; WL--Weak Gravitational Lensing; SK--Satellite Kinematics. Col. (3) 
Mean density of dark matter haloes in units of the critical density of the universe 
($\Delta_{\rm vir}=\bar{\rho}/\rho_{\rm crit}$). Col. (4) Redshift zero cosmological matter density. 
Col. (5) Redshift zero cosmological dark energy density. 
Col. (6) Redshift zero Hubble parameter in units of $\rm km\,s^{-1}\, Mpc^{-2}$. 
Col. (7) Reference for stellar masses and stellar initial mass function: B03--Bell \etal (2003); 
BR07--Blanton \& Roweis (2007); G06--Gallazzi \etal (2006); K03--Kauffmann \etal (2003); 
P07--Panter \etal (2007); diet-Sal.-- Salpeter (1955) $-0.15$dex; Kroupa--Kroupa (2001); 
Chab.--Chabrier (2003); Col (8) Offset we apply to stellar mass, in dex.
}
  \begin{tabular}{lcccccccc}
\hline
\hline  
Reference  &  Method    &  $\Delta_{\rm vir}$ & $\Omega_{\rm m}$ & $\Omega_{\Lambda}$ &
$H_0$ & Stellar Masses & Mass Offset & Morphology\\
(1) &(2) & (3) & (4) & (5) & (6) & (7) & (8) & (9)\\
\hline
Yang \etal 2007       & GC & \pt71.4\pt& 0.238   & 0.762   & 73 & B03, diet-Sal. & $-0.1$ & All\\
Moster \etal 2010     & AM & \pt95.62  & 0.26\po & 0.74\po & 72 & P07, Chab. & $\phantom{+}0.0$ & All\\
Guo \etal 2010        & AM & 200.0\pt& 0.25\po & 0.75\po & 73 & BR07, Chab. & $+0.1$ & All\\
Behroozi \etal 2010   & AM & \pt91.8\pt& 0.27\po & 0.73\po & 70 & BR07, Chab. & $+0.1$ & All\\
Mandelbaum \etal 2006 & WL & \pt54.0\pt& 0.30\po & 0.70\po & 70 & K03, Kroupa & $\phantom{+}0.0$ & Elliptical/Disk\\
Mandelbaum \etal 2008 & WL & \pt54.0\pt& 0.30\po & 0.70\po & 70 & G06, Chab. & $\phantom{+}0.0$ & Elliptical \\
Schulz \etal 2010     & WL & \pt60.0\pt& 0.30\po & 0.70\po & 70 & K03, Kroupa & $\phantom{+}0.0$ & Elliptical\\
Conroy \etal 2007     & SK & 200.0& 0.30\po & 0.70\po & 100h& B03, diet-Sal. & $-0.1$ & All/Red/Blue\\
More \etal 2010       & SK & \pt47.6\pt& 0.238   & 0.762   & 73 & B03, diet-Sal. & $-0.1$ & All/Red/Blue\\
Klypin \etal 2010     & SK & \pt91.8\pt& 0.27\po & 0.73\po & 70 & BR07, Chab. & $+0.1$  & All\\
\hline
\hline
\end{tabular}
\\
\label{tab:mm}
\end{table*}

\section{Observations}
\label{sec:obs}
This section gives an overview of the observational data we use to
derive the relation between optical and virial velocities of galaxies.
This paper primarily uses observational data from studies based on the
Sloan Digital Sky Survey (SDSS), Data Release 4.  The SDSS (York \etal
2000; Stoughton \etal 2002; Abazajian \etal 2004; Adelman-McCarthy
\etal 2006) is an extensive photometric and spectroscopic survey of
the local universe.

\subsection{Virial Masses}
\label{sec:mm}
There are a number of techniques that are used to determine virial
masses\footnote{Note that when we refer to the virial mass of a dark
  matter halo we implicitly include the baryonic mass as well as the
  dark matter mass} of dark matter haloes. Direct observational
measurements of virial masses can be obtained with weak galaxy-galaxy
lensing (e.g., Brainerd \etal 1996; Hudson \etal 1998; Wilson \etal
2001; Guzik \& Seljak 2002; Hoekstra, Yee \& Gladders 2004; Mandelbaum
\etal 2006; Cacciato \etal 2009) and satellite kinematics (e.g.,
Zaritsky \& White 1994; Prada \etal 2003; van den Bosch \etal 2004;
Conroy \etal 2005). A limitation of these techniques is that the
signal for individual galaxies is usually too weak to give
statistically significant measurements. This means that halo masses
must be obtained by stacking many galaxies together (typically within
some luminosity or stellar mass bin).  Another technique, known as the
(sub) halo abundance matching method (Kravtsov \etal 2004; Vale \&
Ostriker 2004; Conroy \etal 2006; Conroy \& Wechsler 2009), determines
the relation between stellar mass and halo mass by assuming that there
is a one-to-one mapping (sometimes scatter in this mapping is
included) between the number density of dark matter haloes (including
sub-haloes) and the number density of observed galaxies as a function
of stellar mass.  A related method, which we term the group catalog
method, is to run a galaxy group finder and to assign each group a
halo mass using abundance matching (this time not including
sub-haloes) under the assumption that there is a one-to-one mapping
between halo mass and the total stellar mass of all group members.

In this paper we consider measurements based on all of these
techniques: Weak lensing (WL) from Mandelbaum \etal (2006; 2008) and
Schulz \etal (2010); Satellite kinematics (SK) from Conroy \etal
(2007); More \etal (2010); and Klypin, Prada \& Montero-Dorta
(2010, in prep); halo abundance matching (AM) from 
Moster \etal (2010), Guo \etal (2010) and Behroozi \etal
(2010); Group catalogs from Yang \etal (2007). The SK measurements
from Conroy \etal (2007) are somewhat different from what is reported
in that paper. Here, we use stellar masses based on Bell \etal (2003),
and we use halo masses derived from samples of SDSS galaxies that have
not been randomly diluted in number density (as was done in order to
mimic the DEEP2 selection function).  Otherwise the sample and
methodology is identical to Conroy \etal (2007).

\subsubsection{Homogenization}
All of these methods are cosmology dependent to some extent. Deriving
halo masses from weak galaxy-galaxy lensing or satellite kinematics
requires some knowledge of the structure of the dark matter halo. The
halo abundance matching method requires a halo mass function (which
depends directly on a number of cosmological parameters).

The virial masses were calculated under different assumptions and
cosmologies. See Table~\ref{tab:mm} for the values of virial radius
overdensity ($\Deltavir$), redshift zero matter density ($\Omega_{\rm
  m}$), redshift zero dark energy density ($\Omega_{\Lambda}$),
redshift zero Hubble parameter ($H_0$). Note that for Yang \etal
(2007) the halo masses are based on the halo mass function of Warren
\etal (2006). The conversion between this mass function and spherical
overdensity masses is non-trivial, as the $\Deltavir$ is dependent on
resolution. Here we assume that the Warren \etal (2006) halo mass
function corresponds to $\Deltavir=300\rho_{\rm mean}$.  While it is
not possible to bring the measurements from each author onto exactly
the same cosmology, we can at least adopt the same definition of halo
mass.  For the results in this paper, the choice of halo definition is
arbitrary, so we adopt the most convenient definition.

Halo masses are commonly defined via $\Deltavir$, the overdensity of
the halo with respect to the critical density of the Universe:
\begin{equation}
\label{eq:vir}
\langle \rho \rangle
\equiv M_{\rm vir} / \frac{4}{3} \pi R_{\rm vir}^3 = \Deltavir
\,\rhocrit.
\end{equation}
Here $M_{\rm vir}$ is the virial mass, $R_{\rm vir}$ is the virial
radius, and $\rhocrit = 3 H_0^2/8\pi G$ is the critical density of the
Universe. 

In this paper we adopt $\Deltavir=200$, so that halo masses are given
by $M_{200}$, halo sizes by $R_{200}$, and halo virial velocities by
$V_{200}$.  This definition results in a simple conversion between
halo virial size, virial velocity, and virial mass at redshift $z=0$:
\begin{equation}
\label{eq:vmr}
  \log \frac{V_{200}}{[\kms]} = \log \frac{R_{200}}{[\hkpc]} = \frac{1}{3}\log \left(G \frac{M_{200}}{[\hMsun]}\right),
\end{equation}
where $G\simeq4.301\times 10^{-6} (\kms)^2 \,\kpc \,\Msun^{-1}$.  This
definition also has the advantage that halo masses are independent of
the matter density of the universe (which is still subject to a
significant uncertainty).

The conversion between the halo masses defined with different values
of $\Deltavir$ depends on the concentration of the halo. For the halo
masses from Mandelbaum \etal (2008) and Schulz \etal (2010) we use the
halo concentrations that these authors derive from the weak lensing
fits. For the other authors we adopt the halo concentration - mass
relation for relaxed haloes in a WMAP 5th year cosmology (WMAP5,
Dunkley \etal 2009), from Macci\'o \etal (2008): $\log_{10} c = 0.830
-0.098 \log_{10} (M_{200}/10^{12}h^{-1}\Msun)$.

\begin{figure*}
\centerline{
\psfig{figure=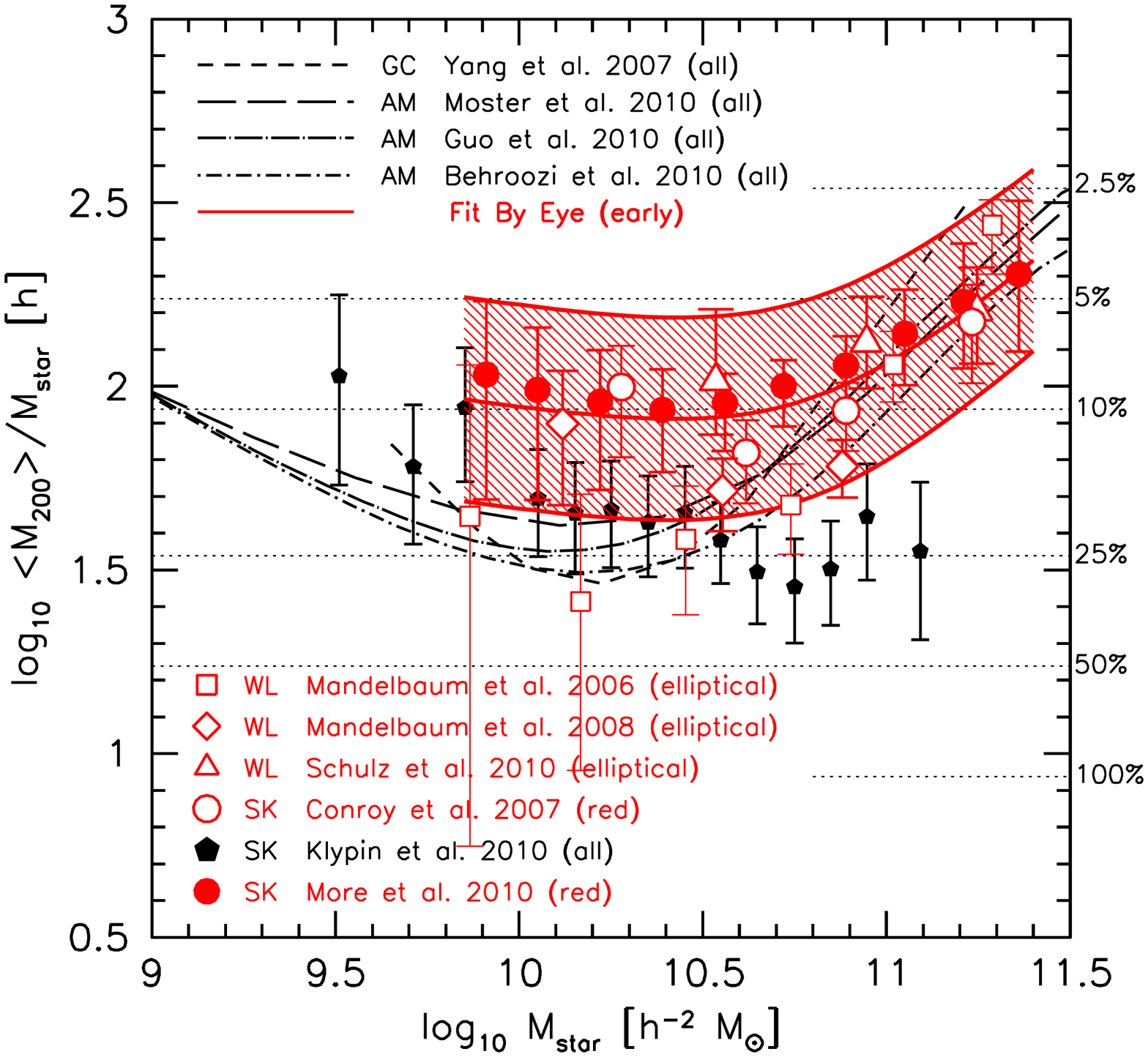,width=0.49\textwidth}
\psfig{figure=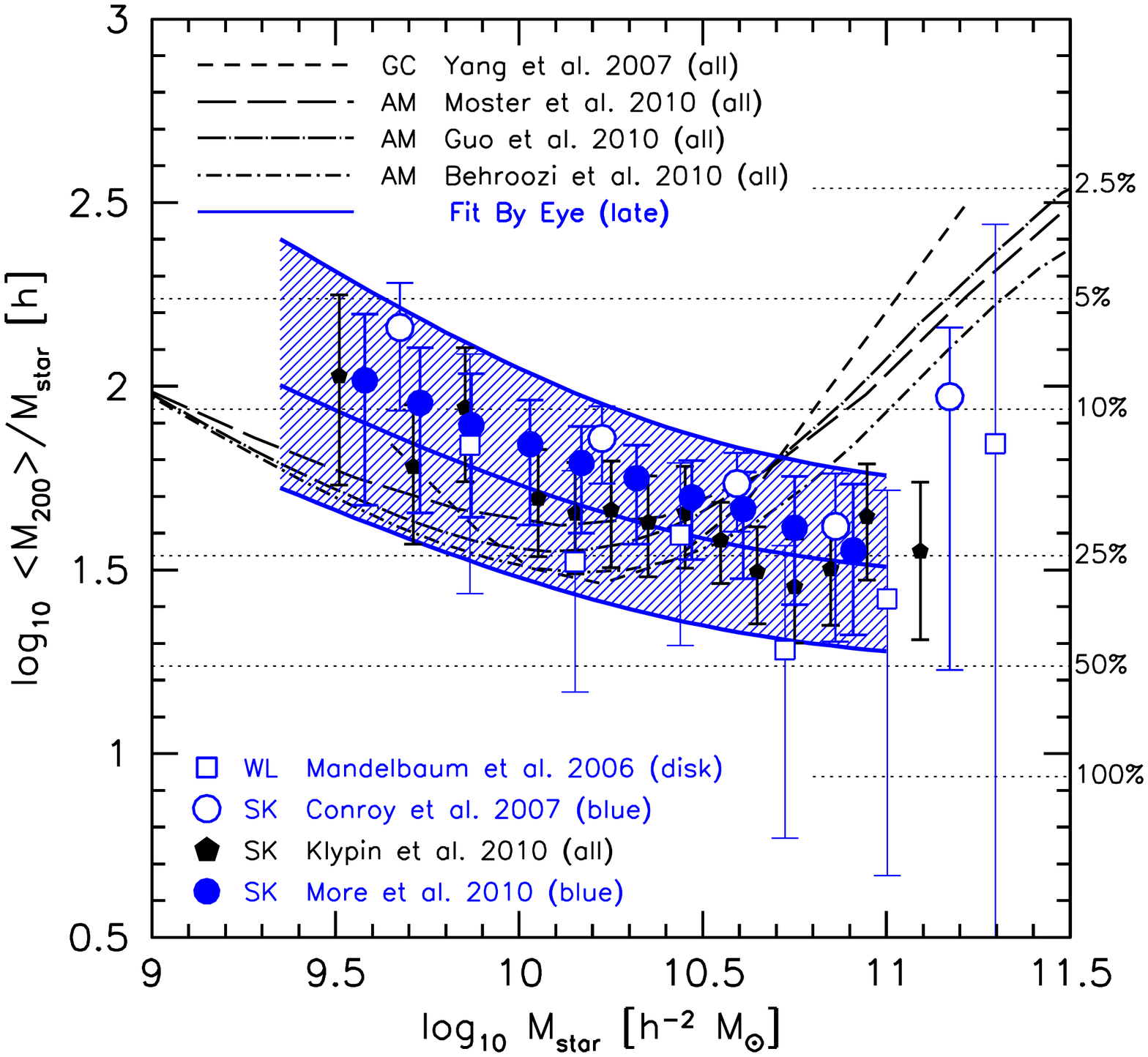,width=0.49\textwidth}}
\caption{ Relation between average virial mass at fixed stellar mass,
  $\langle M_{200}\rangle$, and stellar mass, $M_{\rm star}$,
  expressed in terms of $\langle M_{200}\rangle/M_{\rm star}$ vs
  $M_{\rm star}$, using data from the literature as indicated. The
  various methods (satellite kinematics, SK, weak gravitational
  lensing, WL, halo abundance matching, AM, and group catalogs, GC)
  all yield consistent results within the quoted uncertainties.  All
  error bars are 95\% or $2\sigma$.  The black lines and points show
  the relations derived from taking all types of galaxies and are
  shown in both panels. The red points show the relation derived for
  early-type galaxies (left panel), and the blue points show the
  relation for late-type galaxies (right panel). The solid lines show
  our fitting formula (determined by eye) for the mean and $2\sigma$
  uncertainty in the relation between $\langle M_{200}\rangle(M_{\rm star})$ and
  $M_{\rm star}$ as given by Eq.~\ref{eq:power2}.  The numbers on the
  right hand y-axis of each panel indicate the percentage of the
  cosmological baryons that reside in stellar mass in central galaxies
  (assuming $h=0.7$ and $f_{\rm bar}\equiv \Omega_{\rm b}/\Omega_{\rm
    m}=0.165$).}
\label{fig:mm}
\end{figure*}

Stellar masses were calculated using different IMFs and assumptions
about the star formation histories and metalicities. While we cannot
account for differences in modeling methods, we can attempt to convert
stellar masses to a uniform IMF. As our fiducial stellar masses we
adopt those from Bell \etal (2003) minus 0.1 dex. Unless otherwise
stated, we assume that these masses correspond to a Chabrier (2003)
IMF.  Our adopted conversions between the stellar masses calculated by
various authors to this definition are given in Table~\ref{tab:mm}. We
note that the differences between various stellar mass estimators vary
with stellar mass, even when the same IMF is adopted. For simplicity
we have chosen conversions most appropriate for stellar masses in the
range $\Mstar \simeq 10^{10}-10^{11} \,h^{-2}\Msun$. The differences
between masses from Bell \etal (2003), Kauffmann \etal (2003), and
Blanton \& Roweis (2007) are based on the comparison made in Appendix
A in Li \& White (2009).

For the Mandelbaum \etal (2008) results the halo masses are given
for $r$-band luminosity bins. We convert these luminosities into
stellar masses by using the Faber-Jackson relations from Gallazzi
\etal (2006), which yield the following mean relation between stellar
mass and $r$-band luminosity: $\log_{10} \Mstar = 1.093 \log_{10}
L_{\rm r} -0.573 $.

\subsection{Relation Between Halo Mass and Stellar Mass}
Fig.~\ref{fig:mm} shows the relation between halo virial mass,
$M_{200}$, and galaxy stellar mass, $\Mstar$, as expressed in terms of
$\langle M_{200}\rangle(\Mstar)/\Mstar$ vs $\Mstar$, for the different
methods described in \S\ref{sec:mm}. There are important differences
between the way the relation between halo mass and stellar mass is
usually expressed between the various methods.  The AM method usually
gives the mean of the log of stellar mass as a function of halo mass:
$\langle\log_{10}M_{\rm star}\rangle (M_{200})$. While the WL and SK
masses give the log of the mean halo mass as a function of stellar
mass: $\log_{10}\langle M_{200}\rangle(M_{\rm star})$. For low stellar
masses these measurements give equivalent relations between halo mass
and stellar mass. However, at high masses, these two definitions give
diverging relations between halo mass and stellar mass, with
$\log_{10}\langle M_{200}\rangle (M_{\rm star})$ giving systematically
lower halo masses than $\langle\log_{10}M_{\rm star}\rangle(M_{200})$
(e.g. see Fig.10, Behroozi \etal 2010). This effect is caused by a
combination of the following: scatter in stellar mass at fixed halo
mass; the number density of haloes strongly decreases with increasing
halo mass; and the relation between stellar mass and halo mass is
shallow at high masses (e.g. More, van den Bosch \& Cacciato 2009).
Since the primary goal of this paper is to determine the $V_{\rm
  opt}/V_{200}$ ratio as a function of observables such as $V_{\rm
  opt}$ and $M_{\rm star}$, we thus focus on $\log_{10}\langle
M_{200}\rangle (M_{\rm star})$. For the AM masses (Moster \etal 2010;
Guo \etal 2010; Behroozi \etal 2010) we use $\langle M_{200}
\rangle(\Mstar)$ provided to us by the authors. These have been
derived assuming a scatter of 0.15, 0.20, and 0.15 dex in stellar mass
at fixed halo mass. For the GC masses (Yang \etal 2007) we use the
$\langle M_{200} \rangle -\Mstar$ relation as calculated by More \etal
(2010).

There are different ways of splitting galaxies into the broad
categories of early- and late-type. Some studies use the bimodality
seen in the color-magnitude diagram (with early-types being red,
late-types being blue), while others divide based on the concentration
of the light profile (with early-types being high concentration,
late-types being low concentration). In this work we assume the two
methods are equivalent, although in detail they are unlikely to be
so. For example, red galaxies can be contaminated with dusty spirals
(which are intrinsically blue), and high concentration galaxies can be
contaminated with blue spiral galaxies with significant bulges.  The
black lines and points in Fig.~\ref{fig:mm} show the measurements
derived from taking all types of galaxies (both panels).  The red
points show measurements for early-type galaxies (left panel), while
the blue points show measurements for late-type galaxies (right
panel).

Since late-type galaxies dominate at low masses, and early-type
galaxies dominate at high masses, the relation between $M_{200}$ and
$\Mstar$ for early-type (late-type) galaxies should converge to those
using all galaxy types at high (low) stellar masses. While this is the
case at high masses, at low masses the AM method gives halo masses
that are systematically lower, by a factor of $\simeq 2$, compared to
the SK method.  We attribute this discrepancy to an unknown systematic
uncertainty in either one of or both the AM or SK methods.

We use the following function for $y(x)$ to provide a fitting formula
for the $y=\langle M_{200}\rangle/\Mstar$ vs. $x=\Mstar$ relation for
early- and late-type galaxies:
\begin{equation}
y= y_0\left(\frac{x}{x_0}\right)^{\alpha}\left[\frac{1}{2}+\frac{1}{2}\left(\frac{x}{x_0}\right)^\gamma\right]^{(\beta-\alpha)/\gamma}.
\label{eq:power2}
\end{equation}
Here $\alpha$ is the logarithmic slope at $x \ll x_0$, $\beta$ is the
logarithmic slope at $x \gg x_0$, $x_0$ is the transition scale, $y_0
= y(x_0)$ is the $y$ value at the transition scale, and $\gamma$
controls the sharpness of the transition.

Given the diverse range of measurement techniques and error
estimation, we fit the relation between $\langle M_{200}
\rangle/\Mstar$ and $\Mstar$ by eye. The parameters of our fits are
given in Table ~\ref{tab:mmfit}. The upper and lower limits were
chosen to bracket the error bars of the measurements (which are
$2\sigma$ or 95\% C.I.), as well as taking into account the halo
abundance matching results for low mass late-types. Thus these limits
can be thought of as roughly $2\sigma$ systematic errors.  Note that
we do not include the WL results for low mass early types from
Mandelbaum \etal (2006) in our fits, as these have been superseded
with more recent measurements (Mandelbaum \etal 2008; Schulz \etal
2010). These more recent papers use isolation criteria to select
central galaxies, rather than selecting all galaxies (of a given type)
and then trying to model the weak lensing signal in terms of centrals
and satellites, as was done in Mandelbaum \etal (2006).

\begin{table}
 \centering
 \caption{Parameters of double power-law fitting formula (Eq.~\ref{eq:power2})
   to the $y=\langle M_{200}\rangle/\Mstar$ vs $x=\Mstar$ relations in Fig.~\ref{fig:mm}. Masses are in units of $h=1$.}
  \begin{tabular}{cccccc}
\hline
\hline  
        &  $\alpha$ & $\beta$ & $\log_{10} x_0$ & $\log_{10} y_0$ & $\gamma$\\
\hline
\multicolumn{6}{l}{Early-types: range $\log_{10} M_{\rm star}=9.85-11.4$}\\
  mean       &-0.15 & 0.85 & 10.8 & 1.97 & 2.0 \\
$+2\sigma$   &-0.15 & 0.80 & 10.8 & 2.24 & 2.0\\
$-2\sigma$   &-0.15 & 0.90 & 10.8 & 1.70 & 2.0 \\
\hline
\multicolumn{6}{l}{Late-types: range $\log_{10} M_{\rm star}=9.35-11.0$}\\
  mean      &-0.50 & 0.00 & 10.4 & 1.61 & 1.0 \\
$+2\sigma$  &-0.65 & 0.00 & 10.4 & 1.89 & 1.0\\
$-2\sigma$  &-0.45 & 0.00 & 10.4 & 1.37 & 1.0 \\
\hline
\hline
\label{tab:mmfit}
\end{tabular}
\end{table}

\subsubsection{Galaxy Formation Efficiency}
The numbers on the right vertical axes in Fig.~\ref{fig:mm} show the
percentage of the cosmologically available baryons that end up as
stars in a given galaxy, where we adopt a cosmological baryon fraction
of $f_{\rm bar} = \Omega_{\rm b}/\Omega_{\rm m}=0.165$, and a Hubble
parameter of $h=0.7$.  We refer to this as the integrated star
formation efficiency\footnote{Note that what we term the integrated
  star formation efficiency should not be confused with the integral
  of the star formation history of a galaxy, which will be larger than
  $\epsilon_{\rm SF}$ due to mass lost through stellar winds and
  supernovae.}, $\epsilon_{\rm SF} = \Mstar/(f_{\rm
  bar}M_{200})$. When including the cold gas, this parameter becomes
the galaxy formation efficiency $\epsilon_{\rm GF} \equiv
(\Mstar+M_{\rm gas})/(f_{\rm bar}M_{200})$. For early-type galaxies we
assume that the cold gas fractions are small enough to be ignored, and
thus $\epsilon_{\rm GF} = \epsilon_{\rm SF}$.

For late-type galaxies the integrated star formation efficiency
increases with increasing stellar mass. Ranging from $\epsilon_{\rm
  SF}=8.9^{+12.0}_{-5.1}\%$ ($2\sigma$) at a stellar mass of
$\Mstar=10^{9.4} h^{-2}\Msun$, to $\epsilon_{\rm
  SF}=26.3^{+18.3}_{-10.8}\%$ ($2\sigma$) at a stellar mass of
$\Mstar=10^{11} h^{-2}\Msun$.  Using the relation between cold gas
fraction (both atomic and molecular) and stellar mass from Dutton \&
van den Bosch (2009): $f_{\rm gas} = 0.374
-0.162(\log_{10}\Mstar-10)$, the corresponding galaxy formation
efficiencies are $\epsilon_{\rm GF}=16.7^{+22.6}_{-9.6}\%$ ($2\sigma$),
$\epsilon_{\rm GF}=33.3^{+23.2}_{-13.7}\%$ ($2\sigma$).

For early-type galaxies the galaxy formation efficiency peaks at
$\epsilon_{\rm GF}=12.4^{+10.1}_{-5.6}\%$ ($2\sigma$), at a stellar mass
of $\Mstar=10^{10.5} h^{-2}\Msun$.  Below this mass the galaxy
formation efficiency is consistent with being constant. Above this
mass the galaxy formation efficiency decreases, such that at the
highest stellar masses probed, $\Mstar=10^{11.4} h^{-2}\Msun$,
$\epsilon_{\rm SF}\simeq 2.8^{+3.9}_{-1.8}\%$ ($2\sigma$).  Note that
this galaxy formation efficiency is only counting the central galaxy.
These massive early-types are centrals in clusters, and so there is a
substantial (possibly dominant) contribution to the total stellar mass
budget in those haloes from satellite galaxies (e.g. Lin \& Mohr 2004).

Using weak galaxy-galaxy lensing derived halo masses Hoekstra \etal
(2005) found an average galaxy formation efficiency of $\simeq 33\%$
for blue galaxies (taking into account the cold gas fraction) and
$14\%$ for red galaxies. These results are consistent with those
presented here. However, we note that since the galaxy formation
efficiencies vary with stellar mass, the average value will depend on
the sample selection, and it is therefore not very meaningful to quote
a single value.

\subsection{The Tully-Fisher Relation} 
The original Tully-Fisher relation was between $B$-band luminosity and
\hi\, linewidth. The term TF relation is currently used to describe
the broader class of relations between luminosity (at optical to
near-IR wavelengths) or mass (stellar or baryonic) and rotation
velocity (measured in a number of different ways) for late-type
galaxies. As such there is no universal, or correct, definition of the
TF relation. Often different versions are better suited to different
applications.

Since the measurements of halo mass that we use in this paper are as a
function of stellar mass, here we focus on the stellar mass TF
relation. For rotation velocities our favoured definition is
$V_{2.2}$, the rotation velocity measured at 2.2 I-band exponential
disk scale lengths. We adopt this for several reasons: The main reason
is that the velocity is measured at a well defined radius, which
enables mass models to be fitted to the TF relation; Secondly, this
radius is small enough that baryons and dark matter contribute roughly
equally to the circular velocity, and thus $V_{2.2}$ provides a
velocity that is sensitive to the distribution of baryons and dark
matter in galaxies.  Lastly, $V_{2.2}$ is relatively easy to measure
observationally from low to high redshifts, and thus the
$V_{2.2}-\Mstar$ relation can be measured for large statistically
complete samples of galaxies.

\begin{figure}
\centerline{
\psfig{figure=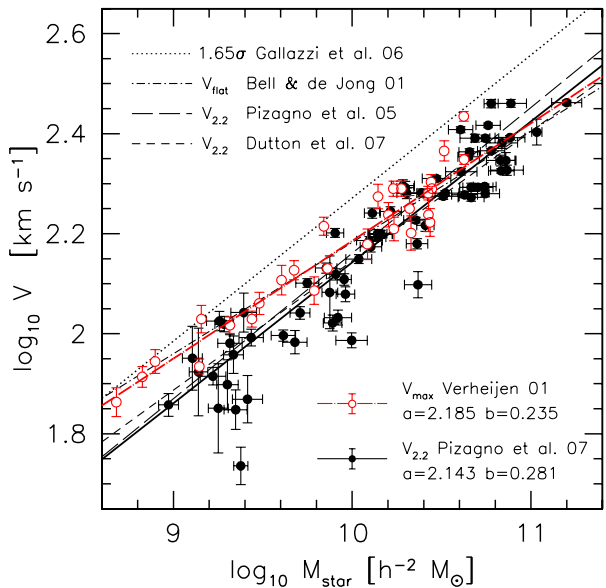,width=0.47\textwidth}}
\caption{ Stellar mass - Tully-Fisher relations derived from the
  Pizagno \etal (2007) and Verheijen (2001) data. The thick lines show
  fits of the form $y=a +b(x-10)$, with best fit $a$ and $b$ as
  indicated. The TF relations from Bell \& de Jong (2001), Pizagno
  \etal (2005) and Dutton \etal (2007) are given by the
  dotted-short-dashed, long dashed, and short dashed lines,
  respectively. For comparison purposes, the dotted line shows the
  Faber-Jackson relation from Gallazzi \etal (2006), after correcting
  velocity dispersions to circular velocities assuming
  $\Vcirc(R_{50})=1.65 \sigma(R_{50})$. }
\label{fig:tf}
\end{figure}

Other velocity measures, such as $V_{\rm max}$, the maximum observed
rotation velocity, or $V_{\rm flat}$, the rotation velocity in the
flat part of the rotation curve, may appear simpler or more
fundamental, but since there is no radius associated with the rotation
velocity, these definitions are not as useful for mass models.
Furthermore, since $V_{\rm flat}$ is measured at large radius, it is
insensitive to the distribution of baryons and dark matter on small
scales, and is observationally challenging to measure.

In this paper we base our stellar mass TF relation on the sample of
$\sim 160$ SDSS galaxies from Pizagno \etal (2007; hereafter P07).
These galaxies were not selected to be late-type, although the
requirement that they have sufficiently strong and extended H$\alpha$
emission to yield a useful rotation curve is equivalent to a late-type
selection based on color.  We compute stellar masses for the P07
galaxies using the $r$-band mass-to-light ratio vs $g-r$ color
relation from Bell \etal (2003), with a -0.10 dex offset
(corresponding to a Chabrier IMF): $\log_{10} \Mstar/L_r = -0.506
+1.097 (g-r)$. In order to minimize inclination uncertainties and
extinction corrections we restrict our sample to galaxies with axis
ratios $0.3 < b/a < 0.5$. These data are shown as black filled circles
in Fig.~\ref{fig:tf}.  The stellar mass TF relation that we derive
here from the P07 data is shown as the black line in Fig.~\ref{fig:tf}
and is given by:
\begin{equation}
\label{eq:tfm}
\log_{10} \frac{V_{2.2}}{[\kms]} = 2.143+ 0.281 \left(\log_{10} \frac{\Mstar}{[10^{10} h^{-2}\Msun]}\right).
\end{equation}
The uncertainty on the zero point is 0.027, the uncertainty on the
slope is 0.003, and the intrinsic scatter is 0.05 dex in $V_{2.2}$.
For comparison, the $V_{2.2}-\Mstar$ TF relation derived in Dutton
\etal (2007), using the much larger but more heterogeneous sample from
Courteau \etal (2007), is shown as the black short dashed line, and
given by:
\begin{equation}
  \log_{10} \frac{V_{2.2}}{[\kms]} = 2.145  +0.259\left(\log_{10} \frac{\Mstar}{[10^{10} h^{-2} \Msun]}\right).
\end{equation}
The uncertainty on the zero point is 0.039, the uncertainty on the
slope is 0.004.  This relation has a shallower slope, but
Fig.~\ref{fig:tf} shows that the differences between these two TF
relations are very small.  Fig.~\ref{fig:tf} also shows the
$V_{2.2}-\Mstar$ TF relation from Pizagno \etal (2005) which was based
on a subset of 81 disk dominated galaxies.  This relation has a
similar slope and zero-point to the one derived here.

The red open circles show the $V_{\rm max}-\Mstar$ relation derived
from data in Verheijen (2001). Stellar masses are calculated using the
relation between $I$-band stellar mass-to-light ratio and $(B-R)$ color
from Bell \etal (2003), with a 0.1 dex offset (corresponding to a
Chabrier (2003) IMF): $\log_{10} \Mstar/L_{I} = -0.505 +0.518
(B-R)$.  We adopt the {\it HST} Key Project distance to the Ursa Major
Cluster of 20.7 Mpc (Sakai \etal 2000), which was also used by Bell \&
de Jong (2001), but larger than the distance of 18.6 Mpc used by
Verheijen (2001). To convert masses to $h^{-2}$ units we adopt
$h=0.7$. A fit to the data is shown by the red dot-long-dashed line
and is given by:
\begin{equation}
  \log_{10} \frac{V_{\rm max}}{[\kms]} = 2.185 + 0.235\left(\log \frac{\Mstar}{[10^{10} h^{-2} \Msun]}\right).
\end{equation}
The uncertainty on the zero point is 0.057, the uncertainty on the
slope is 0.006.  This relation has a shallower slope than the
$V_{2.2}-\Mstar$ relation. At high stellar masses $V_{\rm max}\simeq
V_{2.2}$, and thus the difference is driven by $V_{\rm max}$ being
higher than $V_{2.2}$ for low mass galaxies, which indicates that
rotation curves are still rising at 2.2 disk scale lengths in low mass
($V_{\rm max}\simeq 100 \kms$) galaxies.

For completeness we also show (dot-short-dashed line in
Fig.~\ref{fig:tf}) the $V_{\rm flat}-\Mstar$ TF relation from Bell \&
de Jong (2001), which also used data from Verheijen (2001).  We show
the relation using stellar masses derived from $I$-band luminosities,
$(B-R)$ colors, mass dependent extinction corrections, and
$h=0.7$. This relation has a similar slope and zero point to the
$V_{\rm max}-\Mstar$ relation that we derive from the same data set.

\subsection{The Faber-Jackson Relation}
\label{sec:fj}
Gallazzi \etal (2006) measured the FJ relation for early-type galaxies
using SDSS data.  Velocity dispersions were corrected from the SDSS
3.0 arcsec diameter aperture to the projected $r$-band half light
radius, $R_{50}$, by the standard assumption that the radial profile
of the velocity dispersion has a log slope of $-0.04$ (J{\o}rgensen
1999). Note that these corrections are typically of order a percent.
Stellar masses (from Gallazzi \etal 2005) were computed using a
Chabrier IMF, and assuming $h=0.7$.  The stellar mass FJ relation is
given by:
\begin{equation}
\label{eq:fj}
\log \frac{\sigma(R_{50})}{[\kms]} = 2.054  +0.286 \left(\log \frac{\Mstar}{[10^{10} h^{-2}\Msun]}\right).
\end{equation}
The relation is valid for stellar masses in the range $10^{9} \lta
\Mstar \lta 10^{11.5} h^{-2}\Msun $. The uncertainty on the slope is
0.020, and the scatter is 0.071 dex in $\sigma(R_{50})$.

We adopt the conversion factor between velocity dispersion within the
projected half light radius, $\sigma(R_{50})$, and circular velocity
(at the projected half light radius) as derived in Padmanabhan \etal
(2004): $V_{50}\equiv \Vcirc(R_{50})=1.65\, \sigma(R_{50})$.  This has
an uncertainty of about 10\% depending on the anisotropy profile.  For
comparison, Seljak (2002) assumed $\Vcirc(R_{50}) = 1.5\,
\sigma(R_{50})$, whereas Wolf \etal (2009), argue that $M_{1/2}=3
\sigma_{\rm los}^2 r_{1/2}/G $, independent of the anisotropy. Here
$M_{1/2}$ is the mass enclosed within a sphere of radius $r_{1/2}$,
the deprojected half light radius, and $\sigma_{\rm los}$ is the line
of sight velocity dispersion of the system. Recasting this in terms of
circular velocity: $V_{1/2} = \sqrt{3} \sigma_{\rm los} \simeq 1.73
\sigma_{\rm los}$, which is within 5\% of our adopted conversion.

The principle motivation for this conversion is to put rotation
velocities and velocity dispersions on the same scale, so that the
ratio between the optical and virial velocities of early- and
late-type galaxies can be more directly compared. 
The relation between $V_{50}$ and stellar mass for early types is
shown as a dotted line in Fig.~\ref{fig:tf}. This relation has a very
similar slope to that of the $V_{2.2}-\Mstar$ TF relation for
late-types, but has a higher normalization (in velocity) by $\simeq
0.13$ dex.  

Since we are using different definitions for the optical circular
velocity for early and late-types it is worth discussing how fair it
is to compare them directly. Recall that for late-type galaxies we use
circular velocity measured at 2.2 I-band disk scale lengths, whereas
for early-type galaxies we use circular velocity measured at the
projected half light radius. For a bulge-less exponential disk, 65\%
of the light is enclosed within 2.2 disk scale lengths. For a bulge
plus exponential disk, the fraction of enclosed light will typically
be higher.  For an exponential disk the maximum circular velocity
occurs at 2.16 disk scale lengths, or 1.29 half mass radii. For a
Hernquist sphere (which is close to a deprojected deVaucouluers
sphere) the maximum circular velocity occurs at 0.551 projected half
mass radii, or 0.414 3D half mass radii.

Thus relative to the half light radius we are measuring $\Vopt$ at a
slightly larger radius for late-types than early-types, but relative
to the peak circular velocity of the baryons, we are measuring $\Vopt$
at a smaller radius in late-types than early-types.  However, since
both early and late type galaxies have roughly constant circular
velocity profiles in the optical parts of galaxies (e.g. Rubin \etal
1985; Koopmans \etal 2006), the small differences between our velocity
definitions for early and late-types does not introduce a significant
bias in the $\Vopt-\Mstar$ relations.

The different normalizations of the TF and FJ relations has important
consequences for the universality of the $M_{200}-\Mstar$ and
$V_{200}-\Vopt$ relations. It implies that at most one of these two
relations can be universal (i.e. true for all types of
galaxies). Current observations suggest that both the $M_{200}-\Mstar$
(Fig.~\ref{fig:mm}) and $V_{200}-\Vopt$ (see \S\ref{sec:mw} below)
relations are different for early- and late-types. However, given the
systematic uncertainties on halo masses, we cannot rule out the
possibility that either of these relations is universal. A more
definitive answer will require more accurate measurements of halo
masses as a function of galaxy type.


\section{Relation Between Optical and Virial Circular Velocities}

\begin{figure*}
\centerline{
\psfig{figure=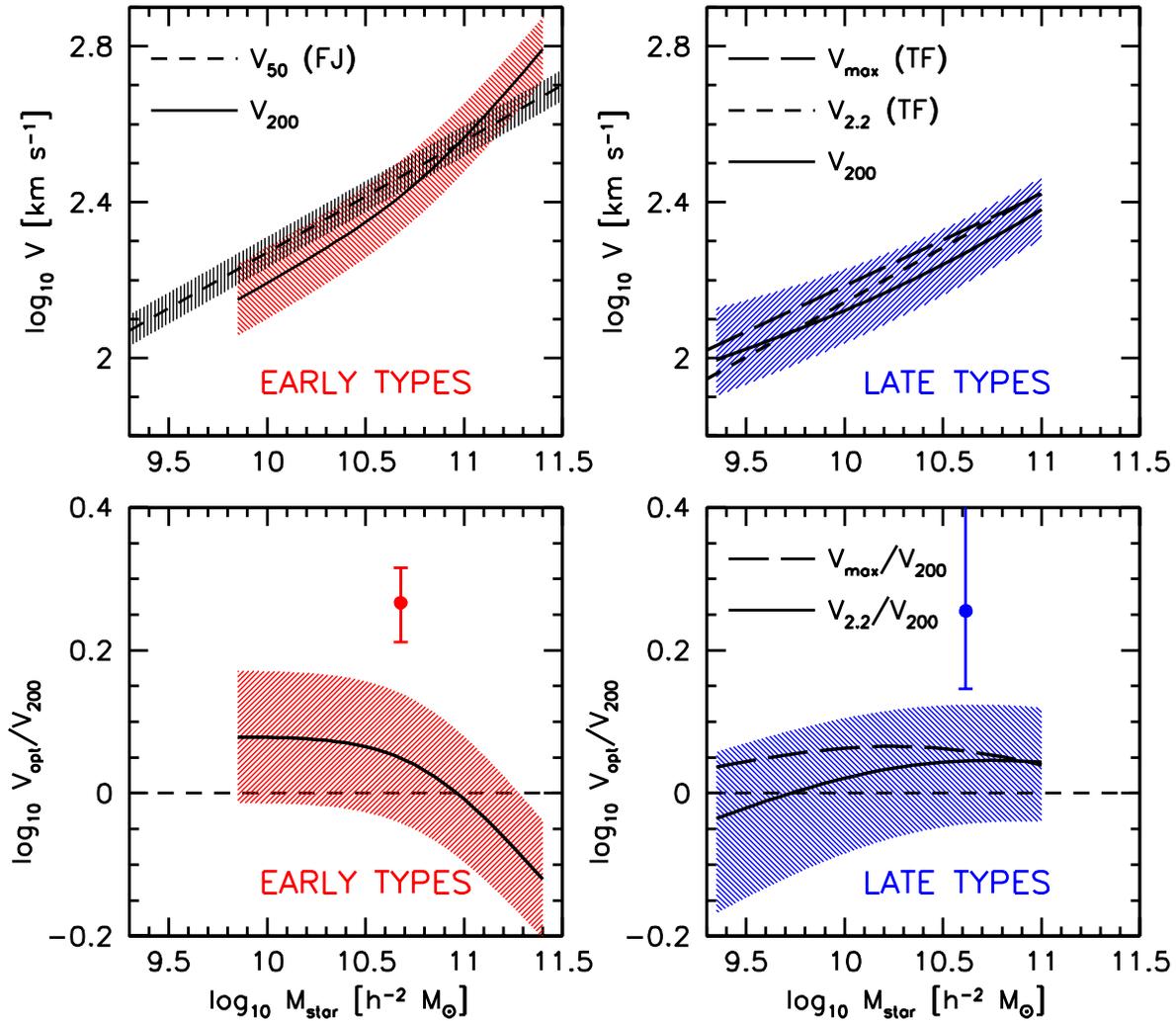,width=0.90\textwidth}}
\caption{Circular velocity vs stellar mass relations (upper panels)
  for early- (left panels) and late-types (right panels). The shaded
  regions correspond to $2\sigma$ uncertainties. The stellar mass
  Tully-Fisher (TF) and Faber-Jackson (FJ) relations are given by the
  dashed lines. For the FJ relation we have converted velocity
  dispersions into circular velocities assuming
  $V_{50}=1.65\,\sigma(R_{50})$. The uncertainty in this conversion is
  given by the black shaded region, and is small compared to the
  uncertainty in halo virial velocities.  The lower panels show the
  difference between the TF and FJ relations and the halo virial
  velocity relations.  The points with error bars ($2\sigma$) show the
  values of $\Vopt/V_{200}$ derived for $L*$ galaxies by Seljak
  (2002).}
\label{fig:vvall}
\end{figure*}

Given a relation between virial mass and stellar mass, we can
trivially compute a virial velocity - stellar mass relation using
Eq.~\ref{eq:vmr}.  Using the fitting formula to the data in
Fig.~\ref{fig:mm}, these relations for early- and late-type galaxies
are shown in the upper panels of Fig.~\ref{fig:vvall}. By comparing
these relations, with the FJ (Eq.~\ref{eq:fj}) and TF
(Eq.~\ref{eq:tfm}) relations (dashed lines in Fig.~\ref{fig:vvall}),
we can derive the average ratio between optical circular velocity
($V_{50}$ for early-types, and $V_{2.2}$ for late-types) and virial
circular velocity (lower panels in Fig.~\ref{fig:vvall}). Note that
the dominant uncertainty in the $\Vopt/V_{200}$ ratio is the
uncertainty in $V_{200}$. Uncertainty in the conversion from $\sigma$
to $\Vopt$, which is just $\simeq 0.04 \rm dex$ (black shaded region
in upper left panel of Fig.~\ref{fig:vvall}), contributes negligibly,
and is therefore ignored in our error analysis.

For this procedure to be valid requires that the galaxies in the
different data samples (e.g. TF vs $M_{200}/\Mstar$) are
representative of the same population, and that the stellar masses are
comparable.  We have verified that we get consistent results for
$\Vopt/V_{200}$ vs. $\Vopt$ when using the velocity - luminosity
relations (i.e. $V_{200}$ - $L_r$ and $\Vopt$ - $L_r$) instead of
velocity - stellar mass relations. Thus the main systematic
uncertainty in our derivation of $\Vopt/V_{200}$ is possible sample
selection differences between TF and halo mass measurements.

For late-type galaxies the virial circular velocity - stellar mass
relation ($V_{200}$ vs $\Mstar$) is shown as blue shaded region in the
top right panel of Fig.~\ref{fig:vvall}. There is some curvature to
the $V_{200} - \Mstar$ relation, but within the uncertainties the data
is consistent with a power-law over the stellar mass range $\Mstar =
10^{9.4}$ - $10^{11.0} h^{-2} \Msun$. The slope and zero point of this
relation closely resembles the optical velocity - stellar mass
relation (dashed lines, Eq.\ref{eq:tfm}). This implies that
$\Vopt/V_{200}\simeq 1$, as is shown in the lower right panel. 

For early-type galaxies the $V_{200} - \Mstar$ relation is not well
described by a single power-law (upper left panel in
Fig.~\ref{fig:vvall}). At low stellar masses the slope is similar to
that of the FJ relation, but at high stellar masses the slope is much
steeper than that of the FJ relation.  This implies that
$V_{50}/V_{200}$ is a strong function of stellar mass (lower left
panel of Fig.~\ref{fig:vvall}). At a stellar mass of $\Mstar \sim
10^{11} h^{-2}\Msun$, $V_{50}\simeq V_{200}$. For higher stellar
masses, $V_{50} < V_{200}$, while for lower stellar masses $V_{50} >
V_{200}$.

\subsection{Comparison with the Literature}

Seljak (2002) used the weak lensing virial mass-to-light ratios from
Guzik \& Seljak (2002) and the same method as used here to derive a
$\Vopt/V_{200}$ ratio for L* galaxies. For late-type L* galaxies
($\Vopt=207 \kms$) Seljak (2002) found $\Vopt/V_{200}=1.8$ with a
$2\sigma$ lower limit of $1.4$, there was no upper limit due to the
lensing masses being consistent with zero.  For early-type L* galaxies
($\sigma=177\kms$) Seljak (2002) found $\Vopt/V_{200} = 1.68 \pm 0.2$
($2\sigma$ uncertainty), assuming $\Vopt=1.5\sigma$. Using our adopted
value of $\Vopt=1.65\sigma$ the result from Seljak (2002) corresponds
to $\Vopt/V_{200}=1.85\pm0.22$. These results are shown as filled
circles and error bars in Fig.~\ref{fig:vvall}.  For both early- and
late-types the results from Seljak (2002) are higher than our values,
with the differences being larger than $2\sigma$. We trace this
discrepancy to the relatively low values of $M_{200}/L$ used by Seljak
(2002). These values are only marginally consistent with weak lensing
results of Mandelbaum \etal (2006) which used a much larger sample
than Guzik \& Seljak (2002), as well as improvements in the modelling
and calibration of the lensing signal. Furthermore, the results from
Mandelbaum \etal (2006) have also been superseded with results from
Mandelbaum \etal (2008) and Schulz \etal (2010) which find  even
higher halo masses.

Seljak (2002) used the high value of $\Vopt/V_{200}$ as evidence for
adiabatic contraction of the dark matter halo. In light of the
significantly lower value of $\Vopt/V_{200}$ that we find here, the
conclusions of Seljak (2002) need to be revisited, which we do so in a
future paper (Dutton \etal in prep).

Eke \etal (2006) used the group abundance matching method with the 2dF
Galaxy Redshift Survey Percolation-Inferred Galaxy Group (2PIGG)
catalogue to derive a relation between halo mass (or equivalently halo
circular velocity) and total $b_J$-band luminosity. They then convert
total group luminosity into average central galaxy luminosity. By
comparing their halo circular velocity - central galaxy luminosity
relation with a $B$-band TF relation (from Bell \& de Jong 2001) they
derive a relation between $V_{\rm flat}/\Vvir$ as a function of
$b_J$-band luminosity.  Here $V_{\rm flat}$ is the rotation velocity
in the flat part of the rotation curve, and $\Vvir$ is the circular
velocity at the virial radius, defined with $\Deltavir \simeq 100$
(see Eq.~\ref{eq:vir}).  For $b_J$-band luminosities in the range
$L_{b_J}=10^{8.5} - 10^{9.8}h^{-2} \Lsun$, Eke \etal (2006) found
$V_{\rm flat}/\Vvir\simeq 0.9-1.0$. While for more luminous galaxies
$V_{\rm flat}/\Vvir$ declines with increasing luminosity.

A direct comparison between our results and those of Eke \etal (2006)
is complicated by the fact that Eke \etal (2006) derived $V_{\rm
  flat}/\Vvir$ by comparing a $\Vvir-L$ relation for all types of
galaxies with the TF relation for spiral (late-type) galaxies. There
are also differences between velocity definitions, but these are small
($\simeq 5\%$) and mostly cancel out. With these caveats in mind, the
values of $V_{\rm flat}/\Vvir$ found by Eke \etal (2006) are
consistent, albeit on the low side, with our results for late-type
galaxies. The lower values of $V_{\rm flat}/\Vvir$ found by Eke
\etal (2006) can be mostly explained by the fact that they compared an
extinction un-corrected $V_{\rm vir} - L_B$ relation to an extinction
corrected $V_{\rm flat}- L_B$ relation. This inconsistency biases
$V_{\rm flat}/\Vvir$ low by $\simeq 0.07$ dex.

Guo \etal (2010) use the relation between halo mass and stellar mass
from abundance matching to construct a $V_{\rm max,h}-\Mstar$
relation.  They compare this to the $V_{\rm flat}-\Mstar$ relation of
spiral galaxies from Bell \& de Jong (2001) finding that in general
$V_{\rm flat} \gta V_{\rm max,h}$. Thus their implied values of
$V_{\rm flat}/V_{200}$ are slightly higher than we find here, but they
are within the uncertainties on halo masses.  Since we have also used
a TF relation from Bell \& de Jong (2001), the differences between our
results and theirs must originate from differences in the halo mass -
stellar mass relation. These differences can be seen in
Fig.~\ref{fig:mm}. Below a stellar mass of $\Mstar=10^{10}h^{-2}\Msun$
the Guo \etal (2010) relation is offset to lower halo masses, which
explains the higher $V_{\rm flat}/V_{200}$ implied by their results.
At a stellar mass of $\Mstar\simeq 10^{10.5}h^{-2}\Msun$
(corresponding to $M_{200}\simeq 10^{12.0}h^{-1}\Msun$ and
$V_{200}=163\kms$) the relations cross over. For galaxies with $\Vopt
\simeq 220 \kms$ Guo \etal (2010) should have found $\Vopt < \Vmaxh$.
Their result that $V_{\rm opt} \gta \Vmaxh$ was biased by the fact
that the observed TF relation data that they adopted had only three
galaxies with $\Vopt > 200 \kms$, which happened to be biased to high
$\Vopt$ at fixed stellar mass (see Fig.~\ref{fig:tf}). 

Guo \etal (2010) claim that the reasonable agreement between the
observed $V_{\rm flat}-\Mstar$ TF relation and their derived $V_{\rm
  max,h}-\Mstar$ TF relation implies that the \LCDM cosmology does
seem able to reproduce observed luminosity functions and Tully-Fisher
relations simultaneously. As shown below in \S\ref{sec:dm} we agree
with the conclusion that $\Vopt \simeq \Vmaxh$ (for galaxies in haloes
with $V_{200}\simeq 100-300 \kms$). However, the fact that one can
simultaneously reproduce the LF and TF relations in \LCDM if one sets
$\Vopt=V_{\rm max,h}$ has been known from previous studies (e.g. Cole
\etal 2000; Benson \etal 2003; Croton \etal 2006). The problem is that
$\Vopt\simeq V_{\rm max,h}$ is not naturally reproduced by
cosmological simulations of disk galaxy formation (see \S\ref{sec:mw}
below) or analytic models (Dutton \etal 2007). Rather than being a
fundamental challenge for the \LCDM paradigm, the most likely
explanation for this difficulty is that we still lack an adequate
understanding of galaxy formation, and inparticular the response of
dark matter haloes to galaxy formation.

\subsection{Are Early-Type Galaxy Mass Density Profiles Isothermal? }
It is well known that late-type galaxies often have roughly flat
rotation curves over the radii probed by gas kinematics. More recently
Koopmans \etal (2006, 2009) used a joint lensing and dynamics analysis
of a few dozen strong gravitational lenses from the SLACS survey
(Bolton \etal 2006) to measure the slope of the total (i.e. dark and
baryonic) density profile within the half light radius of early-type
galaxies. These authors found that the total density slope was very
close to isothermal.  Furthermore, by combining the results from the
SLACS lenses with those from the LSD survey (Treu \& Koopmans 2002),
Koopmans \etal (2006) find that the total density slope is close to
isothermal out to a redshift $z=1$.

The range of radial scales probed by these studies is small: from 0.2
to 1.3 galaxy half light radii.  It is thus possible that the galaxies
studied by Koopmans \etal were only locally isothermal over the small
range of radii that they could probe. To obtain a measurement of the
total mass density slope over a larger range of radii, Gavazzi \etal
(2007) measured the weak lensing signal from 18 strong lenses from
SLACS, and found that the total mass density profile is roughly
isothermal out to large (few 100 kpc) radii.

The lower left panel of ~\ref{fig:vvall} shows that while constant
circular velocity profiles cannot be ruled out (at the $2\sigma$
level) the majority of early-type galaxies favor $\Vopt \ne V_{200}$.
However, at a circular velocity of $\Vopt \simeq 350 \kms$
(i.e. $\sigma\simeq215 \kms$, or $\Mstar \simeq 10^{11}h^{-2}\Msun$)
the $\Vopt/V_{200}$ ratio is unity, which is a necessary, but not
sufficient, condition for a galaxy mass distribution to be globally
isothermal.  The lenses used in Gavazzi \etal (2007) had velocity
dispersions $200 \lta \sigma \lta 330 \kms$ (i.e. $ \log_{10} \Vopt
\simeq 2.6 \pm 0.1$). Thus our results are consistent with those of
Gavazzi \etal (2007), but suggest that early-type galaxies can only be
globally isothermal (i.e. a constant circular velocity, $V_{\rm
  circ}$, from the optical half light radius, $R_{50}$, to the virial
radius of the host dark matter halo, $R_{200}$), if at all, over a
narrow range of velocity dispersions.  Further support for the
conclusion that early-type galaxies are not globally isothermal comes
from Klypin \& Prada (2009) who studied the kinematics of satellites,
at distances of 50-500 kpc, around isolated red galaxies in the SDSS.

We remind the reader that, while WL and SK analyses {\it assume} NFW
profiles (which are not isothermal) to derive halo masses, the
resulting halo mass is quite insensitive to the actual shape of the
halo mass profile (e.g. van den Bosch \etal 2004; Conroy \etal 2007). This
is also supported by the fact that halo masses from WL and SK analyses are
consistent with those from halo abundance matching, which makes no
assumption about the structure of dark matter haloes.  Furthermore,
measurements of $\Vopt$ are independent from those of $V_{200}$. This
means, for example, that even if we assumed that haloes are isothermal
in deriving halo masses from SK and WL, this would not guarantee that
$\Vopt/V_{200}=1$.  Thus, there is therefore no circularity in our
conclusion that the mass profiles of early-type galaxies, from the
optical half light radius to the virial radius of the halo, are in
general non-isothermal.

\section{Implications for The Black Hole Mass - Dark  Halo Mass Relation}

There is a well studied correlation between the masses of central
super-massive black holes (BH) and the bulge mass or velocity
dispersion: the $M_{\rm BH}-\sigma$ relation (Magorrian \etal 1998;
Ferrarese \& Merritt 2000; Gebhardt \etal 2000; Tremaine \etal 2002).
There have been attempts to extend this relation to one between black
hole mass and dark halo mass (e.g. Ferrarese 2002; Bandara, Crampton,
\& Simard 2009), by assuming $\Vopt=\Vmaxh$ or $\Vopt=V_{200}$.
However, since this conversion is at best applicable as an average for
samples of galaxies, it is currently not possible to determine the
scatter in the black mass - halo mass relation. Here we focus on the
implications of our results for the slope of the relation between
black hole mass and halo mass.

\begin{figure}
\centerline{
\psfig{figure=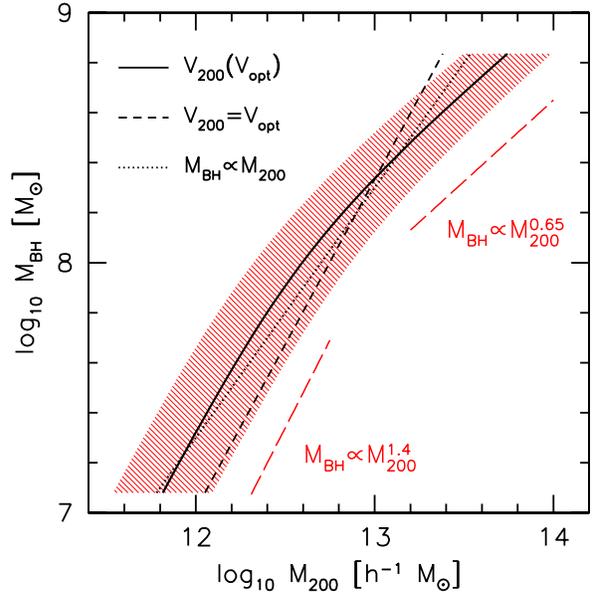,width=0.47\textwidth}}
\caption{Relation between central super-massive black hole mass,
  $M_{\rm BH}$, and halo mass, $M_{200}$. The solid line shows the
  relation derived assuming our relation between $\Vopt$ and $V_{200}$
  for early-types, and the $M_{\rm BH} - \sigma$ relation from
  G\"ultekin \etal (2009). The shaded regions correspond to the
  $2\sigma$ uncertainties in the relation. The solid line has a slope
  of $\simeq 0.65$ at high halo masses, and $\simeq 1.4$ at low halo
  masses, as indicated by the red long-dashed lines. The short-dashed
  line shows the $M_{\rm BH} - M_{200}$ relation derived assuming
  $V_{200}=\Vopt=1.65\sigma$. This relation has a slope of $\simeq
  1.32$. For reference the dotted line shows a linear relation between
  $M_{\rm BH}$ and $M_{200}$. }
\label{fig:mbh}
\end{figure}

By using our relation between halo mass and stellar mass for
early-type galaxies (Eq.~\ref{eq:power2}, Table~\ref{tab:mmfit}),
combined with the FJ relation (Eq.~\ref{eq:fj}) we can derive a
relation between halo mass and galaxy velocity dispersion. By
combining this relation with the $M_{\rm BH}-\sigma$ relation for
elliptical galaxies from G\"ultekin \etal (2009): $\log_{10} M_{\rm
  BH}=(8.23\pm 0.08)+(3.96\pm0.42)\log_{10} (\sigma/200\kms)$, valid
for $3\times 10^6\Msun \lta M_{\rm BH} \lta 3\times 10^9 \Msun$, we
derive a relation between the mean halo mass as a function of black
hole mass. This relation is shown in Fig.~\ref{fig:mbh}, where the
shaded region corresponds to the $2\sigma$ uncertainty in the $M_{200}
- \Mstar$ relation. The parameters of our fitting function are given
in Table~\ref{tab:mbh}.

Observational and theoretical studies of the $M_{\rm BH}-M_{200}$
relation often make the simplifying assumption that the observed
circular velocity (in the optical part of a galaxy) is proportional to
the circular velocity at the virial radius: $\Vopt=\gamma V_{200}$,
where a singular isothermal sphere corresponds to $\gamma=1$. For
example Croton (2009) use this assumption, with $\gamma=1$, to derive
a relation between black hole mass and halo mass $M_{\rm BH} \propto
M_{200}^{1.39\pm 0.22}$.  Bandara \etal (2009) use strong lensing
masses from the SLACS survey (Bolton \etal 2006), and the implicit
assumption of $\gamma=1$, to derive a relation between black hole mass
and halo mass, $M_{\rm BH} \propto M_{200}^{1.55\pm 0.31}$. Adopting
$V_{200}=\Vopt=1.65\sigma$ results in a slope $\simeq 1.32\pm0.14$
(shown by the dashed line in Fig.~\ref{fig:mbh}). The slope of this
relation is consistent with our results for black holes less massive
than about $10^8\Msun$, but for more massive black holes the
assumption of $\Vopt \propto V_{200}$ breaks down.  Shankar \etal
(2006) use the abundance matching method between black holes and dark
matter haloes to determine a relation between black hole mass and halo
mass. They find a relation with a double power-law. At high halo
masses ($\gta 10^{12}\Msun$) the slope is 1.25. This agrees with our
relation at a halo mass of $M_{200} \simeq 10^{12}\hMsun$, but by a
halo mass of $M_{200}=10^{13}\hMsun$ our relation has a significantly
shallower slope of $\simeq 0.65$.

\begin{table}
 \centering
 \caption{Parameters of double power-law fitting formula
   (Eq.~\ref{eq:power2}) to the $y=M_{200} \,[h^{-1}\Msun]$ vs
   $x=M_{\rm BH} \,[\Msun]$ relations in Fig.~\ref{fig:mbh}. }
  \begin{tabular}{cccccc}
\hline
\hline  
        &  $\alpha$ & $\beta$ & $x_0$ & $y_0$ & $\gamma$\\
\hline
\multicolumn{6}{l}{Early-types: range $\log_{10} M_{\rm BH}=7.08- 8.84 \Msun$}\\
  mean       &0.751 & 1.633 & 8.158 & 12.92 & 1.766\\
  $+2\sigma$ &0.751 & 1.589 & 8.158 & 13.19 & 1.766\\
  $-2\sigma$ &0.751 & 1.678 & 8.158 & 12.65 & 1.766 \\
\hline
\hline
\label{tab:mbh}
\end{tabular}
\end{table}

The slope of the $M_{\rm BH}-M_{200}$ relation has theoretical
interest, because different mechanisms for black hole growth are
purported to predict different slopes (e.g., Wyithe \& Loeb 2003).
Our results show that the slope of this relation is not constant,
varying from $\simeq 1.4$ at halo mass of $M_{200} \sim 10^{12}\hMsun$
to $\simeq 0.65$ at halo mass of $M_{200}\sim 10^{13.5}\hMsun$.  This
might imply that different mechanisms for black hole growth are
important at different halo masses. However, since the $M_{\rm
  BH}-\Mstar$ relation is roughly linear, the mass dependent slope of
the $M_{\rm BH}-M_{200}$ relation is driven by the mass dependent
slope of the $\Mstar-M_{200}$ relation.  This latter relation is
determined by the physics of galaxy formation, which might not be
directly governed by the physics of supermassive black holes. Thus the
relation between black hole mass and stellar mass may be more
fundamental (in terms of black hole physics) than the relation between
black hole mass and halo mass.  We note that the change in slope of
the $M_{\rm BH}-M_{200}$ relation occurs at $M_{200} \sim 10^{12.6}$,
corresponding to the mass scale of galaxy groups.

Using a cosmological simulation for the co-evolution of black holes
and galaxies Booth \& Schaye (2010) argue that black hole masses are
determined by the masses of their host dark matter haloes. As
supporting evidence for their conclusion they find good agreement
between the $M_{\rm BH}-M_{200}$ relation in their simulations and the
observational relation from Bandara \etal (2009) (i.e. a slope of
1.55). Our results for the slope of the $M_{\rm BH} - M_{200}$ are
consistent with the prediction of Booth \& Schaye (2010) for black
hole masses between $\simeq 10^7$ and $\simeq 10^8\Msun$, but for
higher black hole masses our results have a much shallower slope.
Thus black hole masses cannot be universally determined by the masses
of their host dark matter haloes via the mechanism proposed by Booth
\& Schaye (2010).


\section{Comparison with $\Lambda$CDM Haloes}
\label{sec:dm}

The maximum circular velocity, $\Vmaxh$, of an NFW (Navarro, Frenk, \&
White 1997) halo is reached at a radius, $r_{\rm max,h} \simeq 2.163
r_{-2} = (2.163/c) R_{200}$, where $r_{-2}$ is the NFW scale
radius\footnote{The NFW scale radius is the radius where the
  logarithmic slope of the density profile, ${\rm d}\ln\rho/{\rm d}\ln
  r=-2$, hence the notation, $r_{-2}$. }, $R_{200}$ is the virial
radius of the halo, and $c\equiv R_{200}/r_{-2}$ is the halo
concentration. For Milky Way mass haloes, $c\simeq 7$, and thus
$r_{\rm max,h}\simeq 0.3 R_{200}$.  The ratio between $\Vmaxh$ and
$V_{200}$ is a function of the halo concentration, and is given by
\begin{equation}
  \frac{\Vmaxh}{V_{200}} \simeq 0.465 \sqrt {c/A(c)},
\end{equation}
where $A(x) = \ln(1+x) - x/(1+x)$. The factor of 0.465 is equal to
$\sqrt{A(x)/x}$ with $x=2.163$. 

\begin{figure}
\centerline{
\psfig{figure=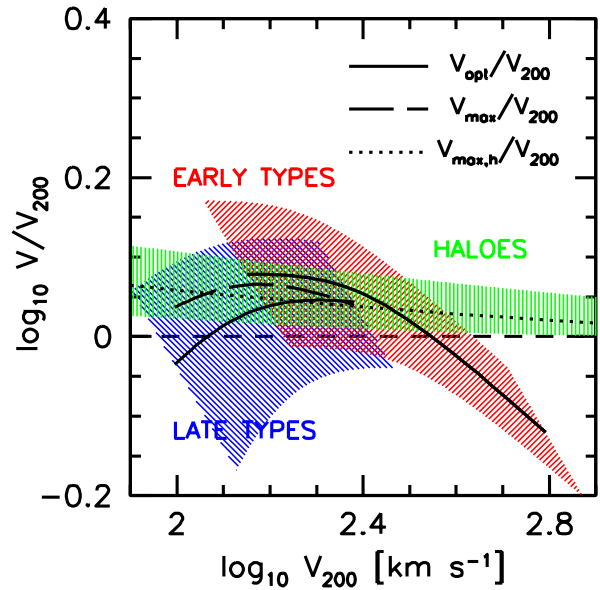,width=0.47\textwidth}}
\caption{ Optical-to-virial velocity ratio vs halo virial
  velocity. The shaded regions correspond to $2\sigma$
  uncertainties. Early-types are shaded red, while late-types are
  shaded blue. For late-types the dotted line shows the relation
  between the maximum rotation velocity and the halo virial
  velocity. The relation between the maximum circular velocity of dark
  matter haloes and the virial velocity of dark matter haloes is given
  by the long dashed line. The green shaded region shows the scatter
  in this ratio due to $2\sigma$ (i.e. 0.22 dex) variation in halo
  concentration.}
\label{fig:vv2}
\end{figure}

Fig.~\ref{fig:vv2} shows the relation between $\Vopt/V_{200}$ and
$V_{200}$ for early- (red shading) and late-type (blue shading)
galaxies.  Note that at a fixed $\Vopt$, uncertainty in $V_{200}$
moves points diagonally from top left to bottom right in the
$\Vopt/V_{200}$ vs $V_{200}$ plot.  The dotted line shows the relation
between $\Vmaxh/V_{200}$ and $V_{200}$ for $\Lambda$CDM dark matter
haloes using the concentration mass relation for relaxed haloes
(Macci{\'o} \etal 2008) in WMAP 5th year cosmology (Dunkley \etal
2009): $\log_{10} c = 0.830 -0.098 \log_{10}
(M_{200}/10^{12}h^{-1}\Msun)$.

The green shaded region shows the variation in $\Vmaxh/V_{200}$
corresponding to $2\sigma$ scatter (0.22 dex) in halo concentrations
(Macci\'o \etal 2008). Note that the uncertainty in the zero point of
the concentration mass relation from uncertainty in cosmological
parameters (mostly $\sigma_8$) is smaller than this.

For late-type galaxies $V_{2.2}/V_{200}$ is consistent with
$\Vmaxh/V_{200}$, but there is a systematic trend for lower
$V_{2.2}/V_{200}$ for low mass haloes.  This trend largely disappears
with $V_{\rm max}/V_{200}$ (long-dashed line), and indicates that the
maximum rotation velocity measured within the optical region of
late-type galaxies is equal to the maximum circular velocity of the
dark matter halo.  Such an assumption is commonly made in
semi-analytic galaxy formation models (e.g. Croton \etal 2006), and
when attempting to infer dark halo masses from optical rotation
velocities (e.g. Blanton \etal 2008; Genel \etal 2008).  Our results
give this assumption empirical justification.

For early-type galaxies the situation is more complex, as
$\Vopt/V_{200}$ is inconsistent with being constant.  For massive
early-types ($V_{200}\gta 250 \kms$), $\Vopt/V_{200}$ decreases with
increasing halo mass, while for lower mass early-types ($130 \lta
V_{200} \lta 250 \kms$), $\Vopt/V_{200} \simeq 1.2$.  For halo
velocities of $V_{200}\simeq 350 \kms$, $\Vopt \simeq V_{200}$.

For the highest halo masses ($V_{200}\simeq 500 \kms$, $M_{200}\simeq
3\times 10^{13} h^{-1}\Msun$), $\Vopt/V_{200} < 1$.  For these halo
masses the halo concentration $c\simeq 5$, and thus the maximum
circular velocity of the halo occurs at $r_{\rm max,h}\simeq 300\, \rm
kpc$. This scale is an order of magnitude higher than the half light
radii of massive early type galaxies, which are of order $10\, \rm
kpc$ (e.g. Shen \etal 2003).  Thus the result that $\Vopt/V_{200} < 1$
can be interpreted as a consequence of the galaxy half light radii
only probing the rising part of the halo circular velocity curve.
This mass scale also corresponds to galaxy groups, so it should not be
a surprise that the baryons in the central galaxy only probe a small
fraction of the virial radius.

For lower halo masses ($130 \lta V_{200} \lta 250 \kms$) there is
evidence that $\Vopt > \Vmaxh$ which would suggest that baryons have
modified the potential well (either by their own gravity, or by
modifying the structure of the dark matter halo through adiabatic
contraction). We will attempt to disentangle these two possibilities
in a future paper (Dutton \etal in prep). However, $\Vopt=\Vmaxh$ is
also consistent with the data (within the $2\sigma$ uncertainties) for
this velocity range.

\begin{figure*}
\centerline{
\psfig{figure=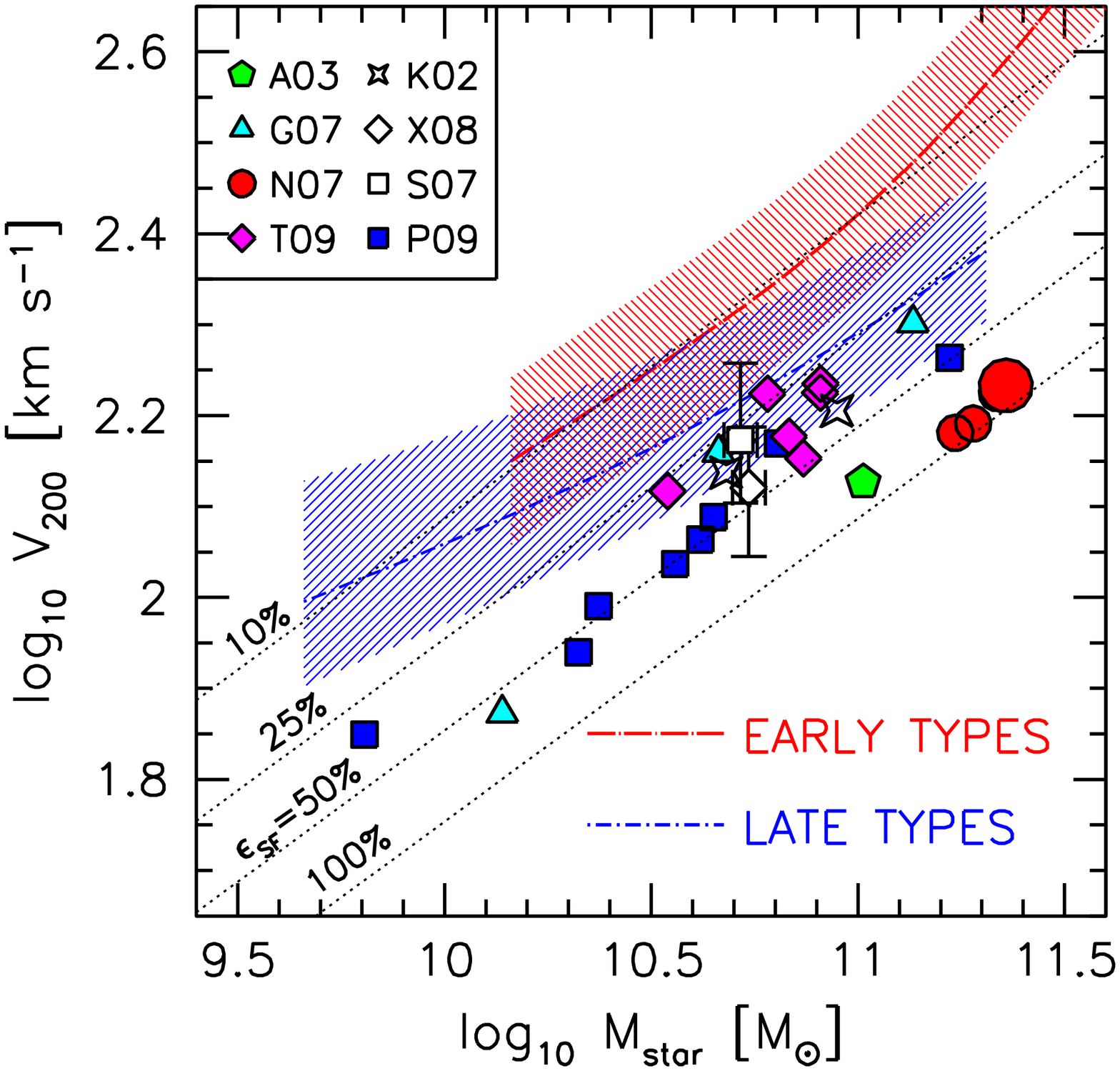,width=0.47\textwidth}
\psfig{figure=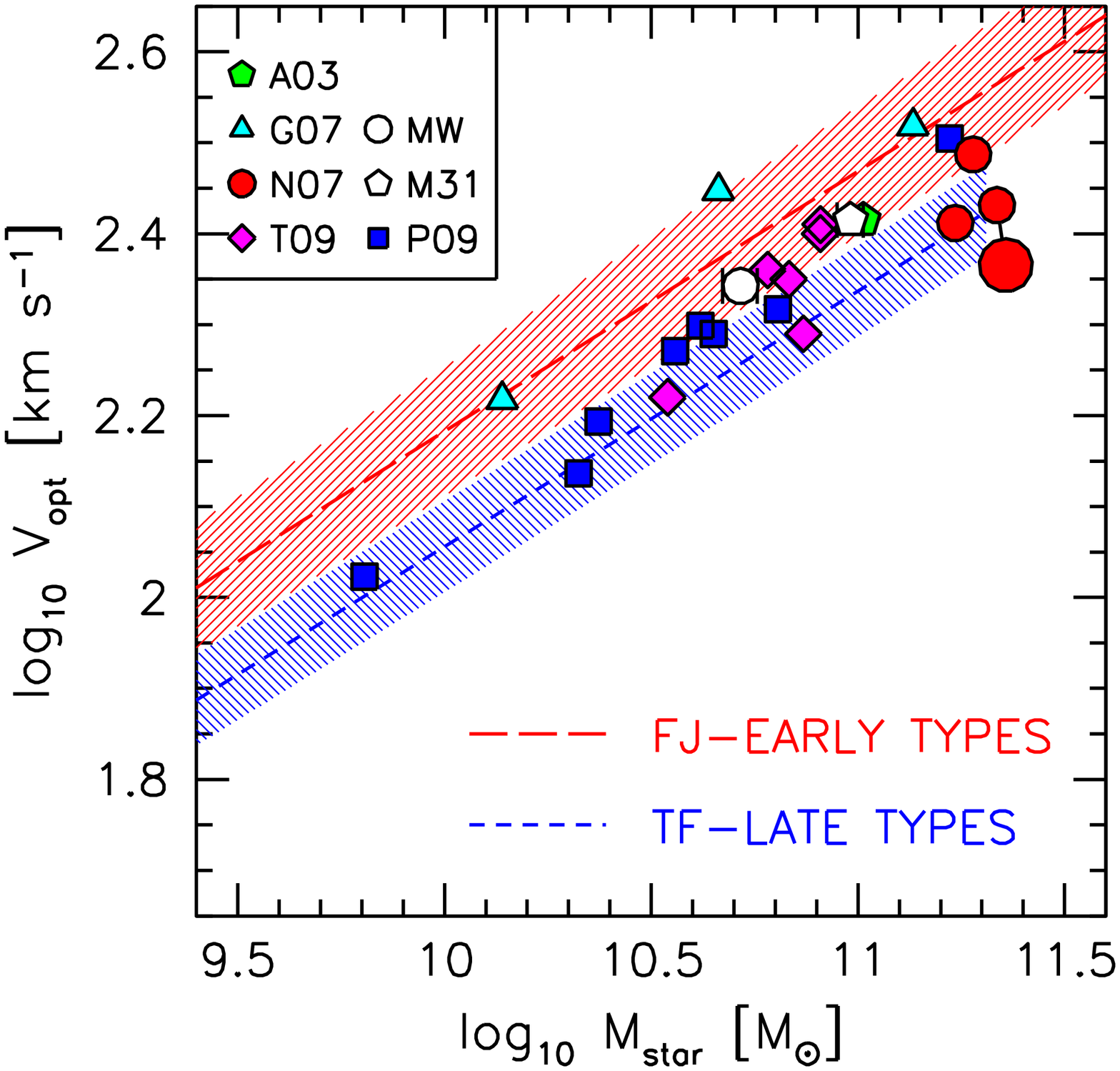,width=0.47\textwidth}
}
\caption{ Comparison between the virial velocity (left) and optical
  velocity (right) TF and FJ relations from this paper (dashed lines
  and shaded regions) and the values for observed (white symbols) and
  simulated (colored symbols) galaxies.  The observed galaxies are M31
  ( Klypin \etal 2002, K02, right star) and the Milky Way (Klypin
  \etal 2002, K02, left star; Smith \etal 2007, S07, white square,
  with 90\% CL error bars; Xue \etal 2008; white diamond, with
  $2\sigma$ error bars, for clarity $\Mstar$ has been shifted by 0.02
  dex). The galaxies from cosmological hydro-dynamical simulations are
  from: Abadi \etal 2003, A03, green pentagon; Governato \etal 2007,
  G07, cyan triangles; Naab \etal 2007, N07, red circles; Tissera
  \etal 2009, T09, magenta diamonds; and Piontek \& Steinmetz 2009,
  blue squares. For the $\Vopt-\Mstar$ relation (right panel) the
  shaded region shows the $1\sigma$ intrinsic scatter, for the
  $V_{200}-\Mstar$ relation (left panel) the shaded region shows the
  $2\sigma$ uncertainty. Stellar masses assume $h=0.7$. In the left
  panel the dotted lines show relations corresponding to integrated
  star formation efficiencies, $\epsilon_{\rm SF}=10,25,50$ \& 100\%,
  assuming a cosmic baryon fraction of 0.165. The MW and M31 galaxies
  are offset from both the $V_{200}-\Mstar$ and $\Vopt-\Mstar$
  relations for late-type galaxies. None of the simulated galaxies
  fall on both the virial and optical TF and FJ relations.}
\label{fig:tffj}
\end{figure*}

\section{Comparison with Observed and Simulated Galaxies}
\label{sec:mw}

\begin{figure}
\centerline{
\psfig{figure=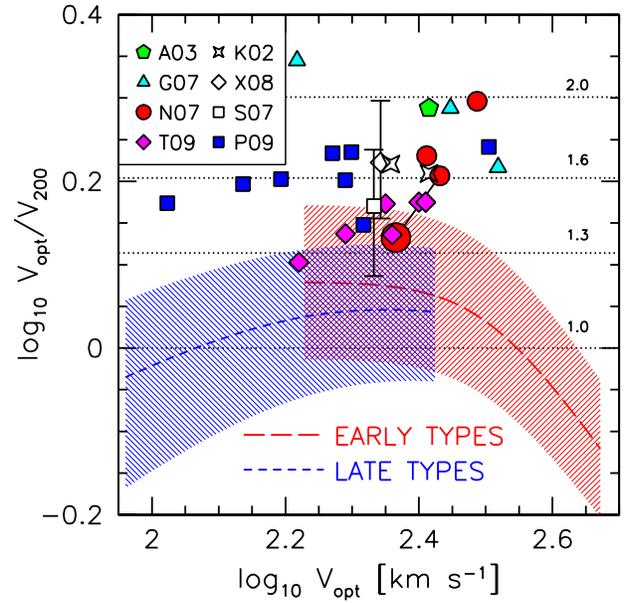,width=0.47\textwidth}
}
\caption{Comparison between the observed $\Vopt/V_{200}$ - $\Vopt$
  relation from this paper (shaded regions correspond to $2\sigma$
  uncertainties) and the values of observed (open symbols) and
  simulated (solid symbols) galaxies. The observed galaxies are M31
  (Klypin \etal 2002, K02, right star) and the Milky Way (Klypin \etal
  2002, K02, left star; Smith \etal 2007, S07, white square, with 90\% CL
  error bars, for clarity $\Vopt$ is shifted by -0.01dex; Xue \etal
  2008; white diamond, with $2\sigma$ error bars). The galaxies from
  cosmological hydro-dynamical simulations are: Abadi \etal 2003,
  A03, green pentagon; Governato \etal 2007, G07, cyan triangles; Naab
  \etal 2007, N07, red circles; Tissera \etal 2009, T09, magenta
  diamonds; and Piontek \& Steinmetz 2009, blue squares. See text for
  further details.}
\label{fig:vvmw}
\end{figure}

Figs.~\ref{fig:tffj} \& \ref{fig:vvmw} shows a comparison between the
$V_{200}-\Mstar$, $\Vopt-\Mstar$ and $\Vopt-V_{200}$ relations that we
derive in this paper (solid lines and shaded regions) with the values
obtained from cosmological simulations (solid symbols) and inferred
for the Milky way (MW) and Andromeda (M31) galaxies (open
symbols). Where necessary we have converted the halo masses and virial
velocities into the $\Deltavir= 200$ definition (see Eq.~\ref{eq:vir})
using the halo concentrations measured or assumed by the authors, or
assuming $M_{200} = 0.83 M_{\rm vir}$ (where $M_{\rm vir}$ corresponds
to $\Deltavir\simeq 100$) where concentrations are not
specified. Table~\ref{tab:vvv} contains the parameters of the
relations between $\Vopt/V_{200}$ and $\Vopt$ that we derive in this
paper, and Table~\ref{tab:vvmw} contains the parameters of the data
points in Figs.~\ref{fig:tffj} \&\ref{fig:vvmw}.

For the MW we adopt $\Vopt=220\kms$ (the IAU value) and a stellar mass
of $5.2\pm0.5\times 10^{10}\Msun$ derived by Widrow, Pym \& Dubinski
(2008) using dynamical models. This mass is consistent with the mass
of $4.8\times 10^{10}\Msun$ from Klypin, Zhao, \& Sommerville (2002)
using dynamical models, the $4.85-5.5\times 10^{10}\Msun$ from Flynn
\etal (2006) using dynamical constraints, and $5\times 10^{10}\Msun$
from Hammer \etal (2007) using stellar population models and a Kroupa
IMF.

For M31 we adopt $\Vopt=260\pm10 \kms$ based on the rotation curve
data compiled by Widrow, Perrett, \& Suyu (2003).  We adopt a stellar
mass of $9.6\pm0.7\times 10^{10}\Msun$. This is consistent with the
mass of $8.9\times 10^{10}\Msun$ favoured by Klypin \etal (2002) using
dynamical models, the $9.5\times 10^{10}\Msun$ from Widrow \etal
(2003) using dynamical models, and the $10.3\times 10^{10}\Msun$ from
Hammer \etal (2007) using stellar population models and a Kroupa IMF.

\subsection{The Milky Way and Andromeda Galaxies}
Applying our results to the Milky Way (MW) ($\Vopt=220 \kms$) and
Andromeda (M31) ($\Vopt=260 \kms$) galaxies we predict that
$\Vopt/V_{200} = 1.11^{+0.22}_{-0.20}$ ($2\sigma$), with corresponding
$V_{200}=198^{+44}_{-32}\kms$ for the MW and $V_{200}=235^{+50}_{-38}\kms$
for M31.  The values of $\Vopt/V_{200}$ that we derive are
significantly lower than suggested by the dynamical models for the MW
and M31 of Klypin, Zhao \& Somerville (2002) which correspond to
$\Vopt/V_{200}=1.66$ (i.e. $V_{200}=137 \kms$) for MW and
$\Vopt/V_{200}=1.62$ (i.e. $V_{200}=161 \kms$) for M31.

Other measurements of the virial mass of the MW also imply relatively
high values for $\Vopt/V_{200}$ (i.e. low values for $V_{200}$):
Using high velocity stars from the RAVE survey Smith \etal (2007) find
$M_{\rm vir}=1.42^{+1.14}_{-0.54} \times 10^{12} \Msun$ which
corresponds to $V_{200}=149^{+32}_{-22} \kms$ (after converting halo
definitions) and $\Vopt/V_{200}= 1.48^{+0.25}_{-0.26}$; Using
kinematics of blue horizontal branch stars in the MW halo, Xue \etal
(2008) find $M_{\rm vir} = 1.0^{+0.6}_{-0.4} \times 10^{12} \Msun
(2\sigma)$ which corresponds to $V_{200}=132^{+22}_{-21} \kms$ (after
converting halo definitions) or $\Vopt/V_{200}=1.67^{+0.31}_{-0.24}$.

The Kahn \& Woltjer (1959) timing argument estimates the mass of the
Local Group from the age of the Universe and the separation and
relative radial velocity of the MW and M31.  Li \& White (2008) use a
cosmological N-body simulation to calibrate the bias and error
distribution of the Timing Argument estimators of the masses of the
Local Group. They derive a combined virial mass of the MW and M31 of
$M_{200}= 5.25^{+4.98}_{-3.30} \times 10^{12}$ (90\% CI).

To interpret this mass we need to divide it between the MW and M31. We
consider two simple ways to do this: Case 1: Equal halo masses.
$M_{200}=2.64^{+2.50}_{-1.72} \times 10^{12}\Msun$, or
$V_{200}=199.1^{+49.7}_{-59.2} \kms$, which implies
$\Vopt/V_{200}=1.10^{+0.47}_{-0.22}$ for the MW, and
$\Vopt/V_{200}=1.31^{+0.55}_{-0.26}$ for M31.  Case 2: Equal
$\Vopt/V_{200}$. This implies a halo mass ratio of $(260/220)^3=1.65$,
and results in $\Vopt/V_{200}=1.21^{+0.51}_{-0.24}$ (90\% CI) for both
the MW and M31.  Thus the timing argument gives $\Vopt/V_{200}$ in
better agreement with our determination, but given the large
measurement uncertainties it is also consistent with
$\Vopt/V_{200}\simeq 1.6$ for the MW.

If the discrepancy between the $\Vopt/V_{200}$ ratios as determined by
our analysis and that measured for the MW and M31 holds under further
study, it would imply that the MW and M31 do not live in typical dark
matter haloes for their optical rotation velocity. This would also
imply that there is substantial scatter in the $\Vopt/V_{200}$ ratio.
Scatter in $\Vopt/V_{200}$ is expected. For example, the analytic
models of Dutton \etal (2007), which are calibrated against the TF and
size-luminosity relations predict a $1\sigma$ scatter of $\simeq 0.05$
dex in this ratio. However, reconciling $\Vopt/V_{200}\simeq 1.6$ with
our measurement of $\Vopt/V_{200} \simeq 1.1$ requires a high sigma
outlier. If we require the MW and M31 to be only slightly atypical, a
larger scatter would be needed, of at least $0.1$ dex. However, a
large scatter in $\Vopt/V_{200}$ may be difficult to reconcile with
the small scatter ($\simeq 0.05$ dex in velocity) in the observed
($\Vopt-\Mstar$) TF relation (Courteau \etal 2007; Pizagno \etal
2007).

If the MW and M31 have atypical $\Vopt/V_{200}$, the left panel of
Fig.~\ref{fig:tffj} shows that the MW and M31 also have atypical
stellar masses for their halo velocities, with the stellar masses
being higher than average. In terms of the $\Vopt-\Mstar$ TF relation,
the MW and M31 are also atypical, being offset to high velocities by
more than $1\sigma$ (in terms of the intrinsic scatter).  The fact
that the MW does not fall on the TF relation (including the $I$-band
luminosity, stellar mass, and baryonic mass variants) has been noted
previously (Flynn \etal 2006; Hammer \etal 2007). Hammer \etal (2007)
found that M31 does fall on the TF relation. This apparent conflict
with our result is due to the different velocity definitions
used. Hammer \etal (2007) adopt $V_{\rm flat}=226\kms$ for M31,
whereas we adopt $V_{2.2}=260\kms$ (based on the rotation curve data
compiled by Widrow \etal 2003).

\begin{table}
 \centering
 \caption{Parameters of double power-law fitting formula (Eq.~\ref{eq:power2})
   to the $y=V_{\rm opt}/V_{200}$ vs $x=V_{\rm opt}$ relations in Fig.~\ref{fig:vvmw}. }
  \begin{tabular}{cccccc}
\hline
\hline  
        &  $\alpha$ & $\beta$ & $x_0$ & $y_0$ & $\gamma$\\
\hline
\multicolumn{6}{l}{Early-types: range$\log_{10} \Vopt=169-470\kms$}\\
  mean       &0.009 & -1.156 & 316.4 & 1.078 & 6.993 \\
  $+2\sigma$ &0.009 & -1.214 & 316.4 & 1.326 & 6.993 \\
  $-2\sigma$ &0.009 & -1.098 & 316.4 & 0.876 & 6.993\\

\hline
\multicolumn{6}{l}{Late-types: range  $\log_{10} \Vopt=91-265\kms$}\\
  mean       &0.407 & -0.186 & 180.1 & 1.099 & 3.559 \\
  $+2\sigma$ &0.348 & -0.186 & 180.1 & 1.321 & 3.559 \\
  $-2\sigma$ &0.585 & -0.186 & 180.1 & 0.886 & 3.559\\
\hline
\hline
\label{tab:vvv}
\end{tabular}
\end{table}

\begin{table*}
 \centering
 \caption{Velocity and mass parameters of the observed and simulated
   galaxies shown shown in Figs.~\ref{fig:tffj} \& \ref{fig:vvmw}}
  \begin{tabular}{llllll}
\hline
\hline  
Galaxy   &  $M_{200}$ & $V_{200}$ & $V_{\rm opt}$ & $V_{\rm opt}/V_{200}$ & $M_{\rm star}$\\
        &  $[10^{11} h^{-1}M_{\odot}]$ & $[{\rm km/s}]$ & $[{\rm
          km/s}]$ &   & $[10^{10} M_{\odot}]$\\
\hline
\multicolumn{5}{l}{This paper}\\
MW      &   $18.0(+14.9,-7.5)$    & $198(+44,-32)$ & 220 &
$1.11(+0.22,-0.20)$ & $5.2(+0.5,-0.5)$\\
M31     &   $30.1(+23.6,-12.4)$    & $235(+50,-38)$ & 260 & $1.11(+0.21,-0.19)$ & $9.6(+0.7,-0.7)$\\
\hline
\multicolumn{5}{l}{Observed: timing argument, Li \& White (2008)}\\
  MW+M31 & $52.5(+49.8,-33.0)$ &- &- &-&- \\
\multicolumn{5}{l}{Observed: halo stars, Xue \etal (2008)}\\
  MW   &  $8.16(+4.90,-3.26)$      & $132(+22,-21)$ & 220 & $1.67(+0.31,-0.24)$&-\\ 
\multicolumn{5}{l}{Observed: high velocity stars, Smith \etal (2007)}\\
  MW   & $11.8(+9.4,-4.5)$      & $149(+32,-22)$ & 220 &
  $1.48(+0.25,-0.26)$& -\\ 
\multicolumn{5}{l}{Observed: dynamical models, Klypin \etal (2002)}\\
  MW   &  8.59 & 137.2 & 228 & 1.66 & 4.8\\
  M31  & 13.7  & 160.5 & 260 & 1.62 & 8.9\\
\hline
\multicolumn{5}{l}{Simulated: Abadi \etal (2003)}\\
       &  5.6 & 134 & 260 & 1.94 & 10.3\\
\hline
\multicolumn{5}{l}{Simulated: Governato \etal (2007)}\\
 DWF1  &  1.38 & 75  & 165 & 2.21 & 1.38\\
 MW1   &  10.0 & 144 & 280 & 1.94 & 4.60\\       
 GAL1  &  26.7 & 200 & 330 & 1.65 & 13.6\\
\hline
\multicolumn{5}{l}{Simulated: Naab \etal (2007)}\\
 A1  &  16.9 & 167.8 & 270 & 1.61 & 21.7\\
 A2  &  17.9 & 171.0 & 232 & 1.36 & 22.8\\       
 C1  &  12.5 & 151.7 & 258 & 1.70 & 17.2\\
 E1  &  13.4 & 155.3 & 307 & 1.98 & 19.0\\
\hline
\multicolumn{5}{l}{Simulated: Tissera \etal (2009)}\\
 Aq-A-5  &  11.0 & 167.9 & 251 & 1.50 & 8.11\\
 Aq-B-5  &   5.2 & 130.8 & 166 & 1.27 & 3.47\\       
 Aq-C-5  &  11.8 & 171.9 & 257 & 1.50 & 8.12\\
 Aq-D-5  &  10.9 & 167.4 & 229 & 1.37 & 6.04\\
 Aq-E-5  &   7.9 & 150.3 & 224 & 1.49 & 6.81\\
 Aq-F-5  &   6.7 & 142.3 & 195 & 1.37 & 7.38\\
\hline
\multicolumn{5}{l}{Simulated: Piontek \& Steinmetz (2009)}\\
 DM-hr1  &  0.82 & 70.7 & 105.4 & 1.49 & 0.64\\
 DM-hr2  &  1.53 & 87.0 & 136.9 & 1.57 & 2.11\\       
 DM-hr3  &  2.18 & 97.8 & 156.1 & 1.60 & 2.36\\
 DM-hr4  &  3.00 & 108.9 & 186.5 & 1.71 & 3.61\\
 DM-hr5  &  3.61 & 115.8 & 199.0 & 1.72 & 4.16\\
 DM-hr6  &  4.28 & 122.5 & 194.8 & 1.59 & 4.48\\
 MW-hr   &  7.48 & 147.6 & 207.5 & 1.41 & 6.41\\
 DM-mr7  &  14.4 & 183.5 & 319.6 & 1.74 & 16.7\\
\hline
\hline
\label{tab:vvmw}
\end{tabular}
\end{table*}

\subsection{Galaxies from Cosmological Simulations}
In Figs.~\ref{fig:tffj} \& \ref{fig:vvmw} we also show the
$V_{200}-\Mstar$, $\Vopt-\Mstar$ and $\Vopt-V_{200}$ relations of
galaxies formed in hydrodynamical cosmological simulations: Abadi
\etal (2003, green pentagon); Governato \etal (2007, cyan triangles);
Naab \etal (2007, red circles); Tissera \etal (2009, magenta
diamonds); and Piontek \& Steinmetz (2009; blue squares). For the
simulations of ``disk'' galaxies of Abadi \etal (2003), and Governato
\etal (2007), we use $\Vopt$ measured at 2.2 disk scale lengths
($I$-band for Abadi \etal 2003, $K$-band for Governato \etal 2007).
These simulations predict $\Vopt/V_{200}\simeq 2$, which is even more
discrepant with our results than the estimates for the MW and
M31. However, as shown by Dutton \& Courteau (2008) the simulated
galaxies from Governato \etal (2007) do not fall on the TF relation
($V_{2.2}$ vs $I$-band luminosity), being offset to high rotation
velocities. This offset is also apparent in the $\Vopt-\Mstar$
relation as shown in Fig.~\ref{fig:tffj}. The cause of this offset is
not clear, but it is likely a combination of insufficient numerical
resolution which leads to artificial angular momentum losses (Kaufmann
\etal 2007), and/or insufficient feedback which results in baryon
fractions that are too high (Dutton \& van den Bosch 2009), and/or too
much adiabatic contraction of the haloes.

The more recent simulations of disk galaxies from Piontek \& Steinmetz
(2009) predict values of $V_{2.2}/V_{200}\simeq 1.6\pm0.1$. Although
these are lower than the $V_{2.2}/V_{200}$ from Abadi \etal (2003) and
Governato \etal (2007), they are still highly inconsistent with our
measurements. Fig.~\ref{fig:tffj} shows that while the simulations
from Piontek \& Steinmetz (2009) fall on the $\Vopt-\Mstar$ relation
at low masses, the simulated galaxies have too much stellar mass at
fixed halo velocity (or halo mass), especially at low masses.

This demonstrates that while cosmological simulations of disk galaxies
have made great progress in producing galaxies that fall on the
$V_{200}-\Mstar$ and $\Vopt-\Mstar$ relations, they still have been
unable to produce galaxies that simultaneously reproduce both of these
relations. Reproducing these relations will provide a key test for
galaxy formation models.

The red circles show the simulations of Naab \etal (2007), with
parameters taken from Johansson \etal (2009). The simulations produced
early-type galaxies (spheroids with no disk component). Three galaxies
have $\Vopt/V_{200} > 1.6$, which is higher (by more than $2\sigma$)
than our derived value.  Galaxy A was re-simulated with 8 times more
particles ($200^3$) resulting in $\Vopt=232 \kms$ and $\Vopt/V_{200} =
1.36$, this galaxy is shown with a larger red circle. This simulation
has $\Vopt/V_{200}$ within $1\sigma$ of our results.  This shows that
numerical resolution is still an important issue for cosmological
simulations that wish to resolve the internal structure of galaxies.
While the simulations from Naab \etal (2007) can produce a relatively
low value of $\Vopt/V_{200}$, Fig.~\ref{fig:tffj} shows that this
galaxies does not fall on the $V_{200}-\Mstar$ or $\Vopt-\Mstar$
relations for early types (red shaded regions).

The magenta diamonds show the simulations of Tissera \etal (2009),
where we have estimated $\Vopt$ from their Figure 9.  These
simulations have $\Vopt/V_{200}\simeq 1.4\pm 0.1$, which is slightly
higher, but consistent within the uncertainties with our results for
early-type galaxies. The simulations of Tissera \etal (2009) have {\it
  only} $10^6$ particles per galaxy, and thus (based on the resolution
tests of Naab \etal 2007) the $\Vopt/V_{200}$ may be biased high by
insufficient numerical resolution. Fig.~\ref{fig:tffj} shows that the
Tissera \etal (2009) simulated galaxies also have stellar masses that
are too high for their halo and optical circular velocities.

The fact that current cosmological simulations produce galaxies with
too many stars has also been shown by Guo \etal (2010). This should
result in an increase in $\Vopt/V_{200}$, and thus it is plausible
that the high values of $\Vopt/V_{200}$ found in the simulations are
just the result of the stellar masses being too high. However, having
the correct amount of stars does not guarantee that a model galaxy has
the correct $\Vopt/V_{200}$. For example, the two most massive
galaxies from Governato \etal (2007) have roughly the correct stellar
masses for their halo masses (for late-types), but they have
$\Vopt/V_{200}\simeq 2$ which is much too high. For a fixed $V_{200}$
and $\Mstar$ the $\Vopt/V_{200}$ ratio also depends on the size of the
galaxy, the halo concentration and on the response of the halo to
galaxy formation (e.g. Mo, Mao, \& White 1998; Dutton \& van den Bosch
2009). Smaller galaxies, higher halo concentrations and halo
contraction all result in higher $\Vopt/V_{200}$. Thus in order to
determine the origin of the $\Vopt/V_{200}$ ratios, the sizes of
galaxies are an essential observational constraint.


\section{Summary}
\label{sec:sum}
We combine measurements of halo virial masses from weak lensing,
satellite kinematics, and halo abundance matching with the
Tully-Fisher (1977) and Faber-Jackson (1976) relations to place
constraints on the {\it average} relation between the optical
($\Vopt$) and virial ($V_{200}$) circular velocities of early-
($\Vopt\equiv V_{50}=1.65 \sigma(R_{50})$) and late-type ($\Vopt\equiv
V_{2.2} = V_{\rm rot}(2.2 R_{\rm d})$) galaxies at redshift $z\simeq
0$.  We summarize our results as follows:

\begin{itemize}

\item The stellar mass to halo virial mass fractions of late-type
  (blue/disk dominated) galaxies increase from $\simeq 0.015$ at a
  stellar mass of $\Mstar = 10^{9.4} h^{-2}\Msun$, to $\simeq 0.043$
  at a stellar mass of $\Mstar \simeq 10^{11.0} h^{-2} \Msun$
  (assuming $h=0.7$). These correspond to integrated star formation
  efficiencies, $\epsilon_{\rm SF} = \Mstar/(f_{\rm bar}M_{200})$, of
  $\epsilon_{\rm SF}\simeq 9\%$ and $\epsilon_{\rm SF}\simeq 26\%$,
  respectively (assuming a cosmic baryon fraction, $f_{\rm
    bar}=0.165$). After accounting for cold gas, the galaxy formation
  efficiencies, $\epsilon_{\rm GF} = (\Mstar+M_{\rm gas})/(f_{\rm bar}
  M_{200})$, are $\epsilon_{\rm GF}\simeq 17\%$ and $\epsilon_{\rm
    GF}\simeq 33\%$, respectively.

\item The central galaxy formation efficiencies of early-type
  (red/bulge dominated) galaxies reach a peak of $\simeq 12\%$ at
  $\Mstar=10^{10.5} h^{-2} \Msun$.  The efficiency drops to
  $\epsilon_{\rm GF}\simeq 2.8\%$ at $\Mstar=10^{11.4} h^{-2}\Msun$.

\item For late-type galaxies $V_{2.2}/V_{200}$ is roughly constant,
  and close to unity, over the range of stellar masses or rotation
  velocities that we probe: ($ 10^{9.35} \lta \Mstar \lta 10^{11}
  h^{-2} \Msun$, or $ 90 \lta V_{2.2} \lta 260 \kms$).

\item For late-type galaxies, the maximum circular velocity of
  $\Lambda$CDM haloes, $\Vmaxh$, is consistent with being equal to the
  maximum observed rotation velocity, $V_{\rm max}$. This result is
  not reproduced by current cosmological simulations of disk galaxy
  formation (e.g. Governato \etal 2007; Piontek \& Steinmetz 2009).

\item Cosmological simulations have been unable to form galaxies that
  fall on both the $V_{200}-\Mstar$ and $\Vopt-\Mstar$
  relations. These relations provide a strong and simple test for
  galaxy formation models.

\item For early-type galaxies $V_{50}/V_{200}$ is not a constant,
  which rules out most early-type galaxies from being globally
  isothermal (i.e. constant circular velocity from the optical half
  light radius to the virial radius of the halo).

\item Early-type galaxies have $V_{50}/V_{200} \simeq 1$ only for
  $\Vopt\simeq350\kms$ (i.e. $\sigma \simeq 215 \kms$). This is
  consistent with the strong plus weak lensing result of Gavazzi \etal
  (2007) which inferred that massive early-types with $\sigma\simeq
  240\kms$ have close to isothermal global density profiles.

\item There is some evidence that for early-type galaxies in lower
  mass haloes ($V_{200} \lta 250\kms$) the $V_{50}/V_{200}$ ratio is
  higher than the $\Vmaxh/V_{200}$ ratio, which would indicate that
  baryons must have modified the potential well, either through their
  own gravity, or by adiabatic contraction of the halo.

\item For early-type galaxies in higher mass haloes ($V_{200} \gta
  500\kms$), $V_{50}/V_{200} < 1 $, indicating that the half light
  radii of the galaxies are much smaller than the NFW scale radius.

\item The mass dependence of the $V_{50}/V_{200}$ ratio for
  early-types implies that the slope of the black hole mass - halo
  mass relation varies with mass.  Specifically it varies from $\simeq
  0.65$ at high halo masses ($\simeq 10^{13.5}\Msun$), to $\simeq 1.4$ at
  low halo masses ($\simeq 10^{12}\Msun$). The shallow slope at high
  mass is at odds with that predicted by the self-regulating feedback
  model of Wyithe \& Loeb (2003) and the cosmological simulation of
  Booth \& Schaye (2010) which both predict a slope of $\simeq 1.5$.

\item For the Milky Way (MW) ($\Vopt=220 \kms$) and Andromeda (M31)
($\Vopt=260 \kms$) galaxies our results predict that 
\begin{equation}
\Vopt/V_{200} = 1.11\, (+0.22,-0.20, 2\sigma)
\end{equation}
This value is significantly lower than those suggested by the dynamical
models for the MW and M31 of Klypin, Zhao \& Somerville (2002) which
correspond to $\Vopt/V_{200}=1.66$ for MW and $\Vopt/V_{200}=1.62$ for
M31.  Other measurements of the virial mass of the MW using kinematics
of halo stars also infer relatively high values: $\Vopt/V_{200}=
1.48^{+0.25}_{-0.26}$ (90\% C.I.)  (Smith \etal 2007);
$\Vopt/V_{200}=1.67^{+0.31}_{-0.24}$ ($2\sigma$) (Xue \etal 2008).  If
both our relation and the dynamical models are correct this
discrepancy would imply that the MW and M31 do not live in typical
dark matter haloes for their optical rotation velocity.
\end{itemize}

\section*{Acknowledgments} 
We thank Benjamin Moster, Qi Guo, and Peter Behroozi for providing
results from their abundance matching.  AAD acknowledges financial
support from the National Science Foundation grant AST-0808133, and
from the Canadian Institute for Theoretical Astrophysics (CITA)
National Fellows program.  CC is supported by the Porter Ogden Jacobus
Fellowship at Princeton University.

This research has made use of NASA's Astrophysics Data System
Bibliographic Services.

Funding for the  Sloan Digital Sky Survey (SDSS)  has been provided by
the Alfred  P. Sloan  Foundation, the Participating  Institutions, the
National  Aeronautics and Space  Administration, the  National Science
Foundation,   the   U.S.    Department   of   Energy,   the   Japanese
Monbukagakusho,  and the  Max Planck  Society.  The SDSS  Web site  is
http://www.sdss.org/.

The SDSS is managed by the Astrophysical Research Consortium (ARC) for
the Participating Institutions. The Participating Institutions are The
University of Chicago, Fermilab, the Institute for Advanced Study, the
Japan Participation  Group, The  Johns Hopkins University,  Los Alamos
National  Laboratory, the  Max-Planck-Institute for  Astronomy (MPIA),
the  Max-Planck-Institute  for Astrophysics  (MPA),  New Mexico  State
University, University of Pittsburgh, Princeton University, the United
States Naval Observatory, and the University of Washington.


\label{lastpage}
\end{document}